\documentclass[traditabstract]{aa}
\usepackage{epsfig,graphicx,natbib,txfonts,url}
\usepackage[usenames]{color}
\bibpunct{(}{)}{;}{a}{}{,}    %% A&A fix to natbib
\def\cite#1{\citealp{#1}}     %% restore old astroncite \cite command

\newcount\longrefs
\def\aap{\ifnum\longrefs=1 {Astron.\ Astrophys.}\else 
                           {A\hbox{\rm \&}A}\fi}
\def\aapr{\ifnum\longrefs=1 {Astron.\ Astrophys.\ Rev.}\else 
                            {A\hbox{\rm \&}AR}\fi}
\def\aaps{\ifnum\longrefs=1 {Astron.\ Astrophys.\ Suppl.}\else 
                            {A\hbox{\rm \&}A Suppl.}\fi}
\def\aj{\ifnum\longrefs=1 {Astron.\ J.}\else 
                          {AJ}\fi} 
\def\ao{\ifnum\longrefs=1 {Applied Optics}\else 
                           {Appl.\ Opt.}\fi} 
\def\aspcs{\ifnum\longrefs=1 {Astron.\ Soc.\ Pacific Conf. Series}\else 
                           {ASP Conf.\ Ser.}\fi} 
\def\apj{\ifnum\longrefs=1 {Astrophys.\ J.}\else 
                           {ApJ}\fi} 
\def\apjl{\ifnum\longrefs=1 {Astrophys.\ J. Lett.}\else 
                            {ApJ}\fi} 
\def\aplett{\ifnum\longrefs=1 {Astrophys.\ J. Lett.}\else 
                            {ApJ}\fi} 
\def\apjs{\ifnum\longrefs=1 {Astrophys.\ J. Suppl.}\else 
                            {ApJS}\fi}
\def\apss{\ifnum\longrefs=1 {Astrophys.\ and Space Science}\else 
                            {Astrophys.\ Space Sci.}\fi}
\def\araa{\ifnum\longrefs=1 {Ann.\ Rev.\ Astron.\ Astrophys.}\else 
                            {ARA\hbox{\rm \&}A}\fi}
\def\azh{\ifnum\longrefs=1 {Astronomicheskii Zhurnal}\else 
                            {Astron.\ Zhur.}\fi}
\def\baas{\ifnum\longrefs=1 {Bull.\ Am.\ Astron.\ Soc.}\else 
                            {BAAS}\fi}
\def\bain{\ifnum\longrefs=1 {Bull.\ Astronom.\ Institutes Netherlands}\else
                            {Bull.\ Astr.\ Inst.\ Neth.}\fi}
\def\gca{\ifnum\longrefs=1 {Geochim.\ Cosmochim.\ Acta}\else 
                           {Geochim.\ Cosmochim.\ Acta}\fi}
\def\grl{\ifnum\longrefs=1 {Geophys.\ Res.\ Lett.}\else 
                           {Geoph.\ Res.\ Lett.}\fi}
\def\iaucirc{\ifnum\longrefs=1 {IAU Circulars}\else 
                          {IAU Circ.}\fi}
\def\ip{\ifnum\longrefs=1 {in press}\else 
                          {in press}\fi}
\def\jgr{\ifnum\longrefs=1 {J.\ Geophys.\ Res.}\else 
                           {J.\ Geophys.\ Res.}\fi}  
\def\jrasc{\ifnum\longrefs=1 {J.\ Royal Astron.\ Soc.\ Canada}\else 
                           {JRAS Can.}\fi}  
\def\mnras{\ifnum\longrefs=1 {Mon.\ Not.\ Roy.\ Astron.\ Soc.}\else 
                             {MNRAS}\fi} 
\def\nat{\ifnum\longrefs=1 {Nature}\else 
                           {Nat}\fi}
\def\pasj{\ifnum\longrefs=1 {Pub.\ Astron.\ Soc.\ Japan}\else 
                            {PASJ}\fi} 
\def\pasp{\ifnum\longrefs=1 {Pub.\ Astron.\ Soc.\ Pacific}\else 
                            {PASP}\fi} 
\def\physscr{\ifnum\longrefs=1 {Physica Scripta}\else 
                            {Phys.\ Scrip.}\fi} 
\def\planss{\ifnum\longrefs=1 {Planetary \& Space Science}\else 
                            {Plan. \& Space Sci.}\fi} 
\def\procspie{\ifnum\longrefs=1 {Proc.\ SPIE}\else 
                            {Proc.\ SPIE}\fi} 
\def\qjras{\ifnum\longrefs=1 {Quarterly J.\ Royal Astron.\ Soc.}\else 
                            {QJRAS}\fi} 
\def\sa{\ifnum\longrefs=1 {Soviet Astron..}\else 
                               {Sov.\ Astron.}\fi}
\def\skytel{\ifnum\longrefs=1 {Sky \& Telescope}\else 
                            {Sky \& Tel.}\fi} 
\def\solphys{\ifnum\longrefs=1 {Solar Phys.}\else 
                               {Sol.\ Phys.}\fi}
\def\ssr{\ifnum\longrefs=1 {Space Science Rev.}\else 
                               {Space\ Sci.\ Rev.}\fi}

\def\nl{,\ } %%\def\nl{\newline}  %% redefine as \newline for mail addresses

\def\ITA{Institute of Theoretical Astrophysics\nl
         University of Oslo\nl
         P.O. Box 1029, Blindern\nl N--0315 Oslo\nl Norway}

\def\OAA{Osservatorio Astrofisico di Arcetri\nl
         Largo Enrico Fermi 5\nl I-50125 Firenze\nl Italy}

\def\SIU{Sterrekundig Instituut\nl Utrecht University\nl Postbus 80\,000\nl
         NL--3508 TA Utrecht\nl The Netherlands}
\def\SPO{NSO/Sacramento Peak\nl P.O. Box 62\nl 
         Sunspot, NM 88349--0062\nl USA}

\def\rmit#1{{\it #1}}              %% italics (RR style, Kluwer)
\def\etal{\rmit{et al.}}           %% use \etal\ for space behind it        
           
\def\ie{\rmit{i.e.,}}              %% , required (Webster 1681)
\def\eg{\rmit{e.g.,}}              %% , required (Webster 1681)
\def\cf{cf.}                       %% no Latin, always Roman (Webster 1686)

\def\specchar#1{\uppercase{#1}}    %% to be redefined for A&A, small caps
\def\CaII{\mbox{Ca\,\specchar{ii}}}

\def\FeI{\mbox{Fe\,\specchar{i}}}

\def\Halpha{\mbox{H\hspace{0.1ex}$\alpha$}} %% \Halpha\ for space behind it

\def\Lyalpha{\mbox{Ly$\hspace{0.2ex}\alpha$}}

\def\CaIIH{\mbox{Ca\,\specchar{ii}\,\,H}}

\def\HK{\mbox{H\,\&\,K}}
\def\KtwoV{\mbox{K$_{2V}$}}

\def\HtwoV{\mbox{H$_{2V}$}}

\def\rmb{{\rm b}}              %% use for indices etc. 

 \def\rmW{{\rm W}}

\def\arcsec{\hbox{$^{\prime\prime}$}}

\def\kms{\hbox{km$\;$s$^{-1}$}}

\def\Mxcm{\hbox{Mx\,cm$^{-2}$}}    %% no 2, damn tex

\def\is{\!=\!}                             %% = in text for tighter spacing
\def\={\hbox{$\!=\!$}}                     %% less space around =

\def\Caline{\mbox{Ca\,\specchar{ii} 854.2~nm}} 
\def\CaIR{\mbox{Ca\,IR}} 
\def\@ #1\par{\par \mbox{}\\ \noindent {\small \tt @ #1}\\[1ex]}  %RR to do 
\def\Mxcm{\hbox{Mx\,cm$^{-2}$}}    %% no 2, damn tex
  
\def\rmit#1{#1}               %% A&A & ApJ: latin abbreviations in Roman
\def\specchar#1{{\sc #1}}     %% A&A style for ionization stages

\begin{document} 

\title{The solar chromosphere at high resolution with IBIS}
\subtitle{IV. Dual-line evidence of heating in chromospheric network}

\titlerunning{Heating in chromospheric network}

\author{G. Cauzzi \inst{1,2}       
        \and
        K. Reardon \inst{1,2}
        \and 
        R.J. Rutten \inst{2,3,4}   
        \and
        A. Tritschler \inst{2}
        \and 
        H. Uitenbroek \inst{2}
}
%RR alphabetic; nice antisymmetry

\authorrunning{G. Cauzzi \etal}

\institute{INAF -- \OAA
           \and \SPO
           \and \SIU
           \and \ITA
}

\date{Received; accepted}
\offprints{G. Cauzzi\\
           \email{gcauzzi@arcetri.astro.it}}

%%%%%%%%%%%%%%%%%%%%%%%%%%%%%%%%%%%%%%%%%%%%%%%%%%%%%%%%%%%%%%%%%% ABSTRACT
\abstract{
  The structure and energy balance of the solar chromosphere remain
  poorly known.  We have used the imaging spectrometer IBIS at the
  Dunn Solar Telescope to obtain fast-cadence, multi-wavelength profile
  sampling of \Halpha\ and \Caline\ over a sizable two-dimensional
  field of view encompassing quiet-Sun network. We provide a first
  inventory of how the quiet chromosphere appears in these two lines
  by comparing basic profile measurements in the form of image
  displays, temporal-average displays, time slices, and pixel-by-pixel
  correlations.  We find that the two lines can be markedly dissimilar in
  their rendering of the chromosphere, but that, nevertheless, both show
  evidence of chromospheric heating, particularly in and around
  network: \Halpha\ in its core width, \Caline\ in its brightness.  We
  discuss venues for improved modeling.  
}


\maketitle

%%%%%%%%%%%%%%%%%%%%%%%%%%%%%%%%%%%%%%%%%%%%%%%%%%%%%%%%%%%%%%%%%%%%%%%%%%%%
\section{Introduction}                              \label{sec:introduction}
%%%%%%%%%%%%%%%%%%%%%%%%%%%%%%%%%%%%%%%%%%%%%%%%%%%%%%%%%%%%%%%%%%%%%%%%%%%%

The solar chromosphere is one of the most difficult solar atmospheric
regimes to observe, model, and understand
 (see recent reviews by
 \cite{1998SSRv...85..187J}; %C Judge+Peter ISSI
 \cite{2006ASPC..354..259J}; %C Judge SacPeak Steinfest
 \cite{2007ASPC..368...49C}; %C Carlsson Coimbra
 \cite{2007ASPC..368...27R}). %C Rutten Coimbra
In this paper we compare spectrally resolved chromospheric image 
sequences taken in the Balmer \Halpha\ line at 656.3~nm and the
\Caline\ line (henceforth abbreviated to \CaIR).  Both lines sample
the chromosphere, but in different manners such that combining them
yields richer diagnostics then when using either line alone.  This has
not been done with the angular, spectral, and temporal resolution and
coverage furnished by the imaging spectroscopy presented here.  In
this initial paper we compare the chromospheric scenes seen in the two
lines by displaying basic spectral measurements across a quiet-sun
field of view. In particular, we show that both lines provide
evidence of network heating, each in its own manner.  So far,
chromospheric heating in or above the magnetic concentrations that
cluster together to constitute the network has been most clearly
evident as brightness enhancement of the cores of
\CaII\ \HK.  A recent discussion of its nature is given by
  \citet{2008ApJ...680.1542H}. % Hasan+AvanB
Additional information from other diagnostics is obviously desirable.

%===========================================================================
%% Fig.~\ref{fig:GONG-evolution} 
%===========================================================================
\begin{figure*}
  \centering \includegraphics[width=180mm]{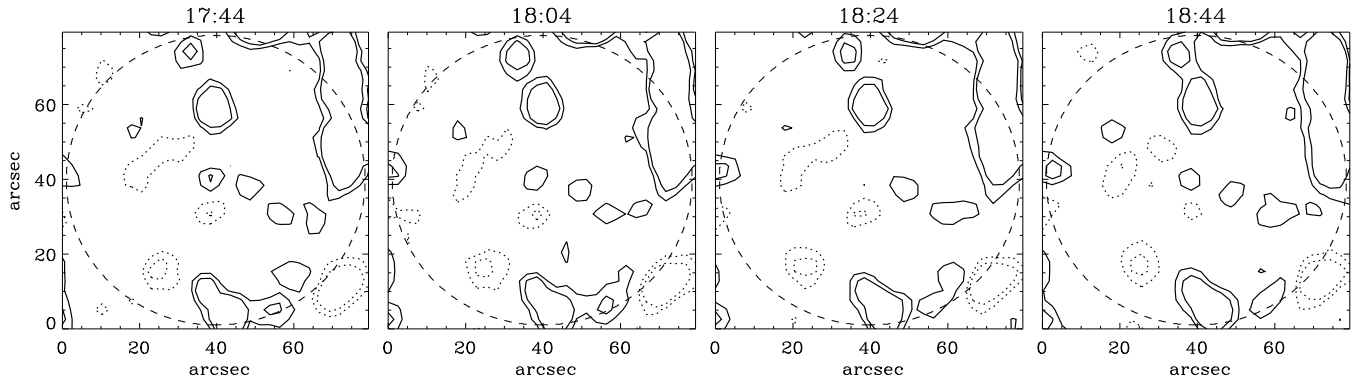}
  \caption[]{
   Magnetic field evolution for the IBIS field of view (dashed
   circle), from GONG data at \url{http://gong.nso.edu}, at 20-minute
   intervals.  The time of observation is specified at the top of each
   panel. The IBIS run was from 17:50 to 18:40~UT. The contours show
   the apparent magnetic flux density at values of 10 and 20~\Mxcm,
%RG Mx cm-2 for observed flux density, Gauss for intrinsic strength
   for positive (solid) and negative (dotted) polarity.
}
\label{fig:GONG-evolution}
\end{figure*}
%===========================================================================

There is an extended literature addressing chromospheric physics using
filtergrams taken in \Halpha\ or \CaII\ \HK, while most work using
\CaIR\ was performed with one-dimensional spectrometry.  
We obtain a fresh view by combining \Halpha\ and \CaIR\ in
%% simultaneous full-field time-sequence sampling, 
imaging spectroscopy,
obtaining high angular resolution with adaptive optics at
the Dunn Solar Telescope (DST) at the U.S. National Solar
Observatory/Sacramento Peak and
%% obtaining 
excellent spectral profile sampling by employing the Interferometric
%RR not full: passband, no outer wings
Bi-dimensional Spectrometer (IBIS) mounted at this telescope.

High-resolution imaging spectroscopy in \CaIR\ was initiated with IBIS
in the study of
  \citet{2007A&A...461L...1V} %C Vecchio++ IBIS 8542 oscs
and continued in the previous papers of this series 
  \citep{2008A&A...480..515C, % Cauzzi++ 8542 IBIS
  2009A&A...494..269V, % Vecchio, Cauzzi & Reardon quiet shocks 8542
  2008arXiv0810.5260R} % Reardon++ K vs 8542
%
%RG I took out Paper I,II,III since not used below
mostly addressing the properties of the quiet chromosphere.  Recent
chromospheric studies with the Swedish 1-m Solar Telescope
combine high cadence with 0.2~arcsec resolution but employed
\Halpha\ only in single-wavelength filtergraph sampling
  (\eg\
   \cite{2006ApJ...647L..73H}; % Hansteen++ dynamic fibrils
   \cite{2006ApJ...648L..67V}; % van Noort++ fast ha
   \cite{2007ApJ...655..624D}; % DePontieu++ dynamic fibrils
   \cite{2007ApJ...660L.169R}). % Rouppe++ QS Ha 
The Dutch Open Telescope has registered many \Halpha\ filtergram
sequences with subarcsecond resolution 
  (\eg\
   \cite{2008SoPh..251..533R}) % RR tomo6
but only with coarse profile sampling and limited in duration by
atmospheric seeing.  Much longer multi-wavelength \Halpha\ filtergram
sequences at this resolution were expected from the tunable filtergraph
onboard {\it Hinode}, but technical problems have severely limited its
ability to sample the \Halpha\ profile at high cadence.
The capability of IBIS and
similar new Fabry-P\'erot instruments (the new G\"ottingen FPI 
at the German Vacuum Tower Telescope 
described by 
\cite{2008A&A...480..265B}, % Bello-Gonzalez+Kneer
and CRISP at the SST 
following the design of
\cite{2006A&A...447.1111S}) % Scharmer FP design 
to supply high-resolution
%full-field 
imaging at fast cadence, with full
profile sampling of both lines, and over fairly long durations
%GC 01120, cut: at which adaptive optics provides good wavefront restoration 
represents a major
step forward in investigations of the solar chromosphere.

Our goal here is to show and compare what the quiet chromosphere looks
like at such improved resolution and coverage of space, time, and
wavelength in the two lines.  Comparison between the two is of
interest because their formation differs intrinsically, as discussed
in Sect.~\ref{sec:sensitivity}. Each has its advantages and disadvantages as
a chromospheric diagnostic.  We limit the analysis here to a
straightforward presentation of three basic profile measurements:
the profile-minimum intensity, the Dopplershift of the profile mimimum, and
the line-core width.  Comparison of the general appearance of the two
lines in these basic properties already yields informative
similarities and differences.

%===========================================================================
%% Table~\ref{tab:IBIS-params}
\begin{table}
\caption{IBIS profile scan parameters.}
\label{tab:IBIS-params}
\centering
\begin{tabular}{lcc}
\hline \hline
  Line & \Halpha  & \CaIR \\
\hline \\[-2ex]
  Nominal wavelength [nm]     & 656.281  & 854.214 \\
  Instrumental profile FWHM [pm]
                              & 2.2      & 4.4     \\
  Wavelength range [nm]       & 0.21     & 0.18    \\ 
  Core sampling interval [nm] & 0.01     & 0.008   \\
  Number of wavelength steps  & 22       & 20      \\
  Profile scan duration [s]   & 5.2      &  4.7    \\
\hline
\end{tabular}
\end{table}
%===========================================================================

%===========================================================================
%% Fig.~\ref{fig:width-measurement}
%===========================================================================

\begin{figure*}
  \sidecaption
  \includegraphics[width=60mm]{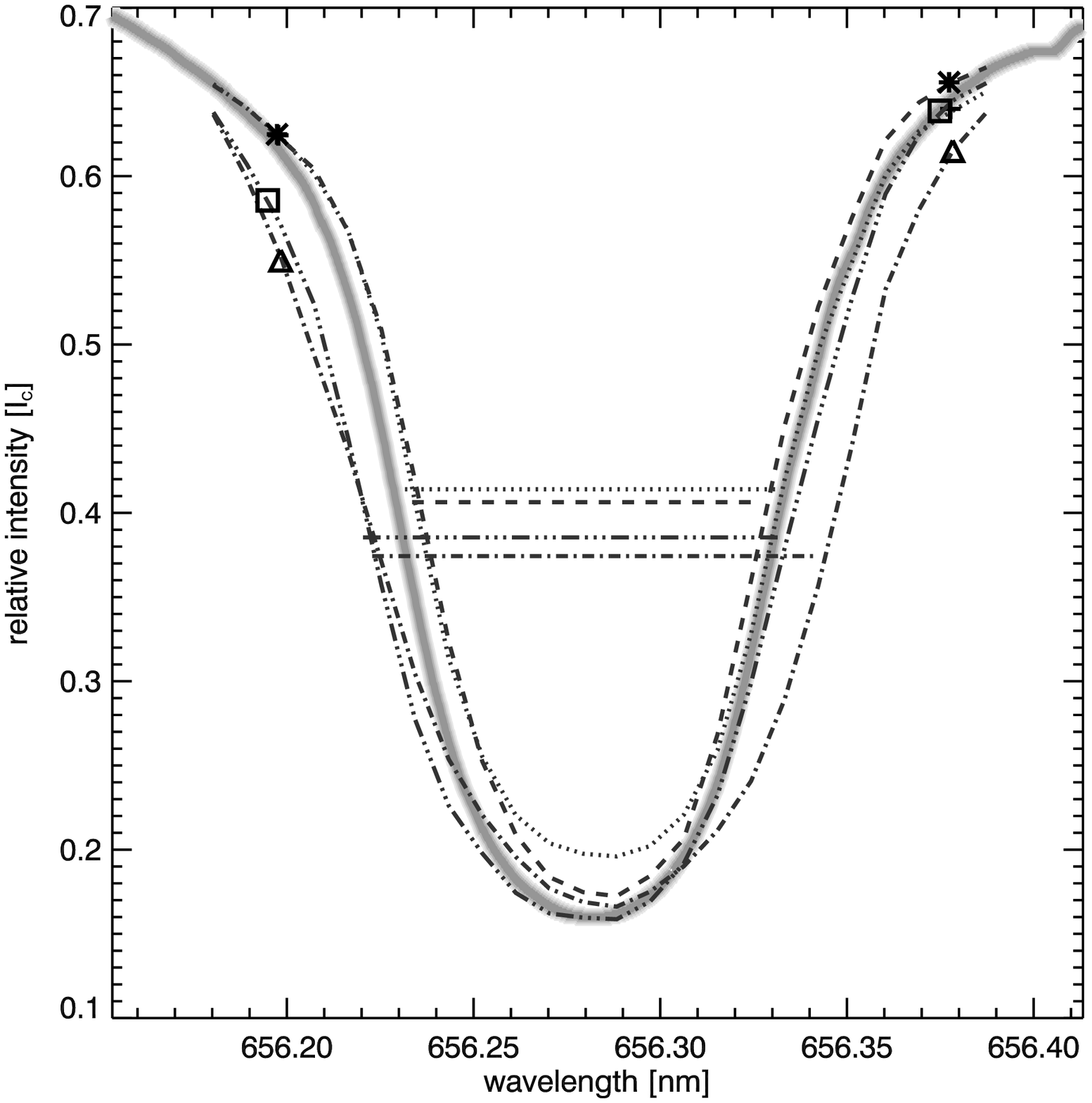}
  \includegraphics[width=60mm]{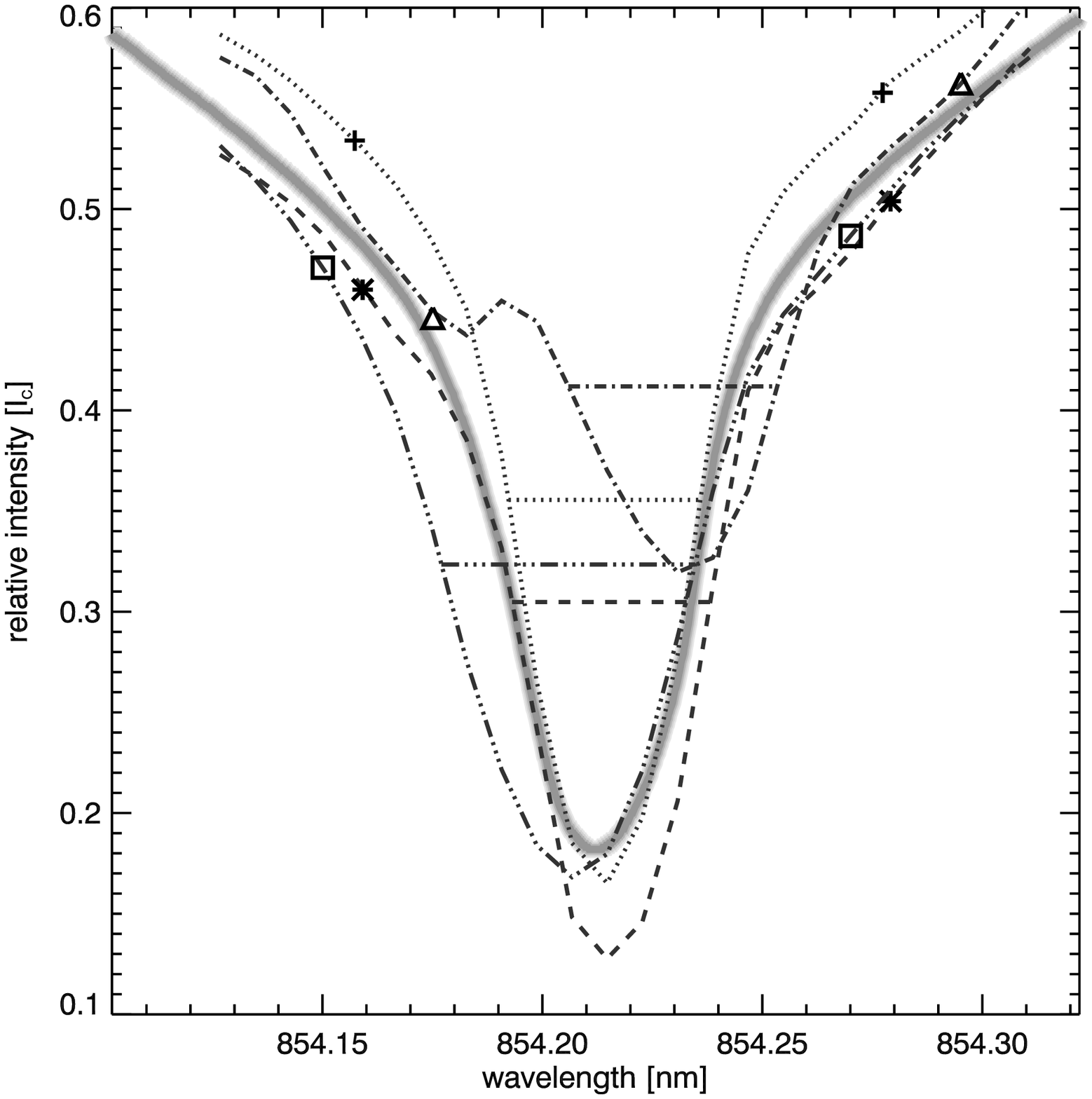}
  \caption[]{
  Core width measurement. {\em Left\/}: \Halpha.
  {\em Right\/}: \CaIR.   Each plot contains profiles
  from four different pixels to illustrate profile variations.  The
  thick grey curve is the solar atlas profile.  The intensity scale is
  in relative units with respect to the atlas continuum value.  For
  each profile $I_\lambda(x,y,t)$ the wing intensities at given
  $\Delta\lambda_l$ from the profile minimum (marked by symbol pairs)
  are averaged.  The core width is the wavelength separation of the
  profile samplings at half the intensity range between the profile
  minimum and this wing average.  The same pixels are selected in
  Fig.~\ref{fig:spectime}.  
}
\label{fig:width-measurement}
\end{figure*}
%===========================================================================

%%%%%%%%%%%%%%%%%%%%%%%%%%%%%%%%%%%%%%%%%%%%%%%%%%%%%%%%%%%%%%%%%%%%%%%%%%%%
\section{Observations, data reduction and profile measurements}  
\label{sec:data+reduction}
%%%%%%%%%%%%%%%%%%%%%%%%%%%%%%%%%%%%%%%%%%%%%%%%%%%%%%%%%%%%%%%%%%%%%%%%%%%%

The imaging spectrometer IBIS is mounted at one of the two high-order
adaptive-optics feeds at the DST.  The latter, described by
  \citet{2004SPIE.5490...34R}, % Rimmele DST AO
serves to stabilize the image and correct wavefront deformations in
real time.
IBIS consists of two Fabry-P\'erot interferometers in tandem and
delivers sequences of narrowband images with the passband stepping
quickly through multiple selectable spectral lines.
%RK avoid ``scan'' 
IBIS is detailed in
  \citet{2006SoPh..236..415C}  % Cavallini, IBIS I
and 
  \citet{2008A&A...481..897R}.  % Reardon+Cavallini, IBIS II 

In this paper we use data taken during 17:50\,--\,18:40~UT on 15 March
2007.  Table~\ref{tab:IBIS-params} specifies the spectral profile
sampling.  IBIS acquired 192 spectral image sets sampling the \Halpha,
\CaIR\ and \FeI~709.0~nm lines with a cadence of 15.4 seconds between
successive scans of all three lines.  The \FeI\ data are not 
considered 
in this paper.
\Halpha\ was scanned with a constant spectral sampling interval of 
0.01~nm, \CaIR\ with 0.008~nm sampling in the core 
and in each wing at three wavelengths separated by 0.016~nm.

The circular field of view has a diameter of 80~arcseconds and was located
near disk center at heliocentric coordinates 3.0~S, 10.1~W degrees,
$\mu \is 0.98$, slightly eastward of a small bipolar
plage.
%RK plage 'equiv ``area'', area of plage is tautology 
Figure~\ref{fig:GONG-evolution} shows the magnetic content and evolution
of the area from full-disk GONG magnetograms.  
It was very quiet, containing a few bipolar patches
of magnetic concentrations that appear as bright points in 
simultaneously taken G-band images.  Only slight topology
changes occurred during the acquired sequence. 

All images were corrected with appropriate dark and flat field
calibrations. The transmission profile of the prefilter was measured
and removed, resulting in good agreement between the 
spatio-temporal averages over the full sequence and the corresponding
profiles in the NSO/FTS solar spectrum atlas of
%
%%  \citet{2007assp.book.....W}. % Wallace+Hinkle+Livingston visatl
%RR bad latex - needs math for cm^-1; so old books.bib item
    \citet[][compare also the average \CaIR\ profile in Cauzzi et al. 2008]{Wallace+Hinkle+Livingston1998}. % visatl 

In order to retrieve the line profile at each pixel, it is necessary
to accurately align the images making up a spectral scan through a line,
including ``rubber-sheet'' destretching to correct remaining
field-dependent deformations by seeing.  The reference for such alignment
was generated from a sequence of broadband images obtained
simultaneously with the narrowband images.  The best broadband image
every 15 seconds (\ie\ during each full scan of all three lines) was
found by measuring granular contrast.  These 192 best broad-band
images per spectral set were then co-aligned to each other, including
destretching through local cross-correlation while taking care not to
remove persistent solar flows. 
The destretched sequence was Fourier filtered in both space and time
using a subsonic filter to remove the Fourier components corresponding
to apparent horizontal motions faster than the photospheric sound
speed (7~\kms).  This resulted in a high-quality master reference
sequence which was then used to evaluate the seeing-induced local
image motion, again using local-correlation tracking, for each
individual broadband image.  The destretch vectors that were derived
for each of these were finally applied to the corresponding simultaneous
narrowband image.  This procedure results in stable, coaligned spectral-scan
image sequences and so in reliable line profiles.  During the moments of
best seeing the resulting image quality is close to the DST
diffraction limit (0.22~arcsec for \Halpha, 0.28~arcsec for \CaIR).

Each individual spectral profile at each pixel in each profile scan was
interpolated onto a uniform wavelength grid with 0.009~nm sampling for
\Halpha\ and 0.008~nm sampling for \CaIR.  
The radially increasing instrumental blueshift, which results from the
placement of the IBIS interferometers in a collimated mount,
%GC 011208, cut:, reaches 0.008~nm and 
was also removed in this resampling.  For each spectral profile we
then determined the wavelength of the profile-minimum by fitting a
second-order polynomial to the five spectral samples centered on the
position with the smallest measured intensity.  The minimum of this fit is taken as
the instantaneous line center, yielding both its intensity (in
relative data units) and its Dopplershift (in \kms\ with respect to
the profile-minimum wavelength of the spatio-temporal profile average
over the flat-field data).
  
The measurement of line width is less straightforward.  The scan
extents specified in Table~\ref{tab:IBIS-params} do not reach the
local continua outside the lines, so that we cannot measure the
profile width at a fixed fractional depth from the continuum.
Furthermore, the shape of these chromospheric lines often varies
significantly, especially in their cores.  For \CaIR\ in particular,
the extended photospheric wings tend to remain relatively stable while
the chromospheric core varies dramatically.  Expressing the latter's
chromospheric properties in terms of photospheric continuum intensity
thus is not a desirable approach.  We therefore use only the inner,
chromospheric part of the profile to define the property we call core
width.

We evaluated these core widths per instantaneous profile per pixel by
first determining and averaging the wing intensities at a specified
wavelength separation $\Delta\lambda_l$ from the measured wavelength of the
profile minimum.  We then defined and measured the core width as the
wavelength separation of the two profile flanks at half the intensity
range between the minimum and this wing intensity.  The separation
parameter was set at $\Delta \lambda_l$\,$=$\,$90$~pm for \Halpha\ and
$\Delta \lambda_l$\,$=$\,$60$~pm for \CaIR. 
These values were selected
through inspecting many profiles with the aim to separate the
chromospheric core from the photospheric wings as well as possible.
The choice remains somewhat arbitrary, but we feel that this procedure
is a better one than halfwidth measurements higher up in the line that
would mix disparate photospheric and chromospheric contributions.
Figure~\ref{fig:width-measurement} illustrates the procedure for the
\Halpha\ and \CaIR\ profiles from four selected pixels at one moment.
Note that the \CaIR\ core varies much more than the \Halpha\ core, and
that the highest \CaIR\ profile has a rather unusual redshift and
blue-wing brightening to which we return below.

Finally, the observed intensities were converted into brightness
temperatures by equating the spatio-temporal average of each line
minimum to the corresponding absolute intensity value in the solar
spectrum atlas of
  \citet{Neckel1999} %T SP, FTS atlases "announcement", not in ADS
  which is a calibrated version of the atlas of
%
%%  \citet{2007assp.book.....W}. % Wallace+Hinkle+Livingston visatl
%RR bad latex - needs math for cm^-1; so old books.bib item
    \citet{Wallace+Hinkle+Livingston1998},
and then applying the inverse of the Planck function.  These formal
temperatures represent kinetic temperatures only when LTE holds for
the source function, which is not the case for either line
(Fig.~\ref{fig:sourcefunctions}).  The conversion serves primarily to
undo the Planck-function sensitivity variation with wavelength.

%===========================================================================
%% Fig.~\ref{fig:instimages}
%===========================================================================
\begin{figure*}
  \centering \includegraphics[width=180mm]{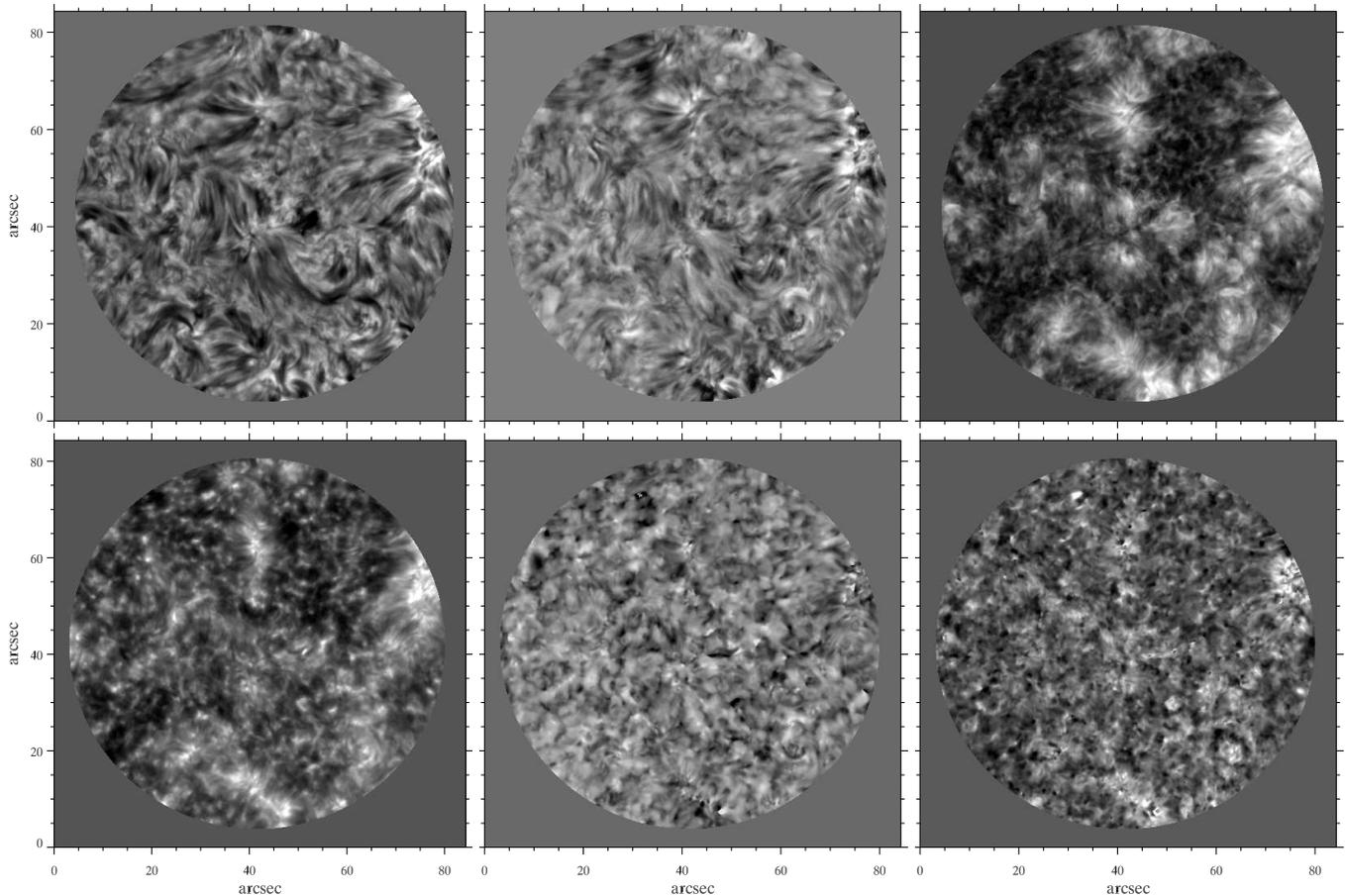}
  \caption[]{
  Image representations of profile measurements at a moment of good
  seeing (scan nr.~11).  
%RR IDL it=10
  Solar North is to the top, West to the right. 
%GC deleted the following sentence
% The IBIS field of view is circular to avoid areas with too much 
% passband shift in wavelength.  
  The greyscale is bytescaled separately for each panel, for the
  Dopplershifts symmetrically with respect to zero (byte value 127;
  blueshift black, redshift white).
  {\em Upper row\/}: \Halpha. {\em Lower row\/}: \CaIR.  {\em First
  column\/}: minimum intensity per pixel, measured as brightness
  temperature.  {\em Second column\/}: Dopplershift of the profile
  minimum per pixel.  {\em Third column\/}: core width,
  measured as in Fig.~\ref{fig:width-measurement}.
}
\label{fig:instimages}
\end{figure*}
%===========================================================================

%===========================================================================
%% Fig.~\ref{fig:meanimages}
%===========================================================================
\begin{figure*}
  \centering \includegraphics[width=180mm]{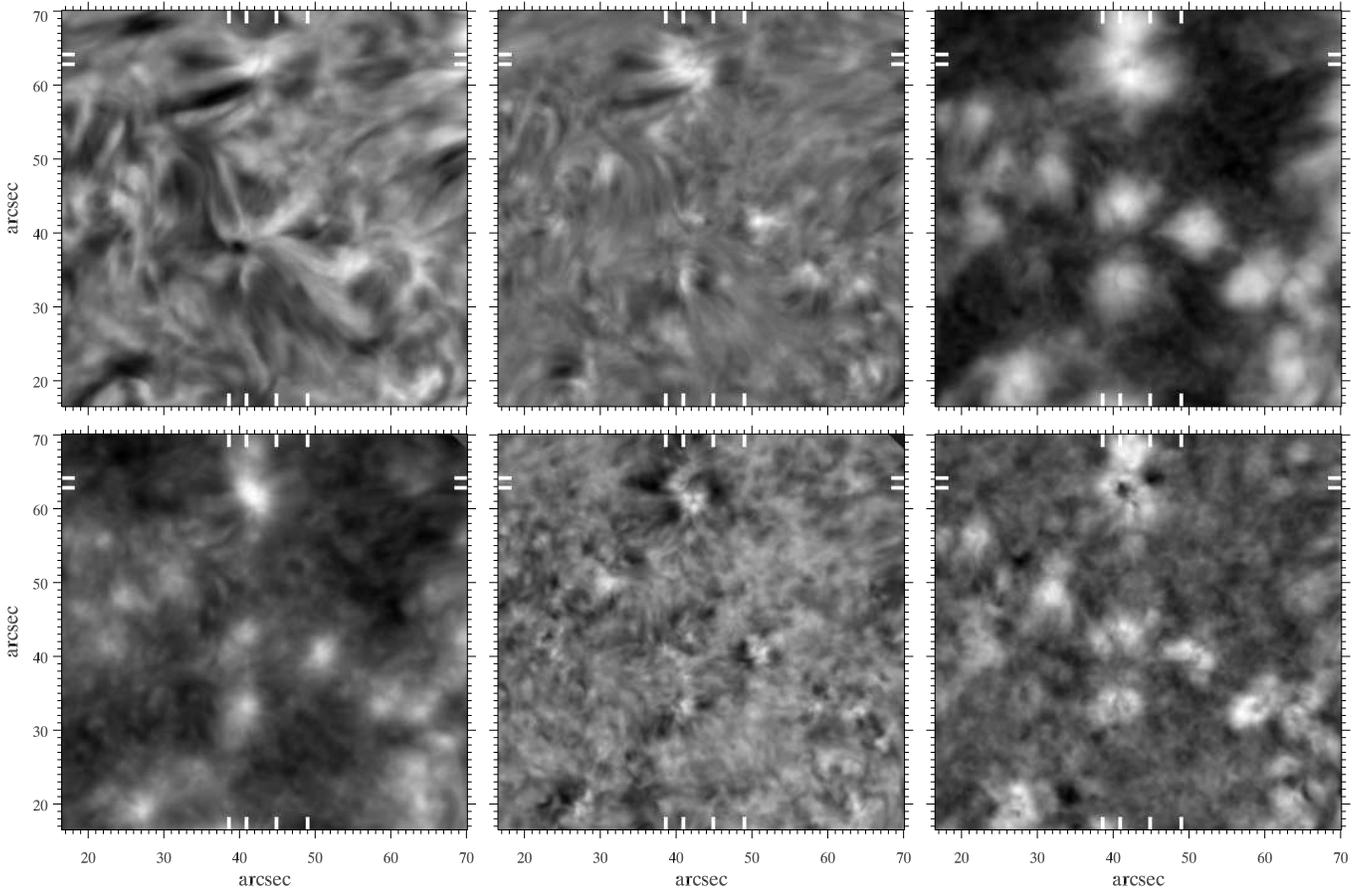}
  \caption[]{
  Temporal averages over the 48-min sequence duration for the central
  part of the field of view, excluding moments of bad seeing (about
  one-third of the profile scans).  The layout furnishes the same
  quantities as in Fig.~\ref{fig:instimages}.  Each panel is
  bytescaled separately, the Dopplershifts symmetrically with respect
  to zero. The two markers along the $y$-axes specify the cut
  locations used in Fig.~\ref{fig:slices}.  They pass through a patch
  of network area in the center and a weaker field concentration near
  $x=63$~arcsec, and through two dark patches of small \CaIR\ core
  width in the last panel.  The four markers along the $x-$ axes
  specify pairs of pixels along each cut for which the spectral
  profile evolution is shown in Fig.~\ref{fig:spectime}.
}
\label{fig:meanimages}
\end{figure*}
%===========================================================================
%RR the weak patch is here at x=62 but shifts quickly to x=64

%===========================================================================
%% Fig.~\ref{fig:slices}
%===========================================================================
\begin{figure*}
  \centering \includegraphics[width=180mm]{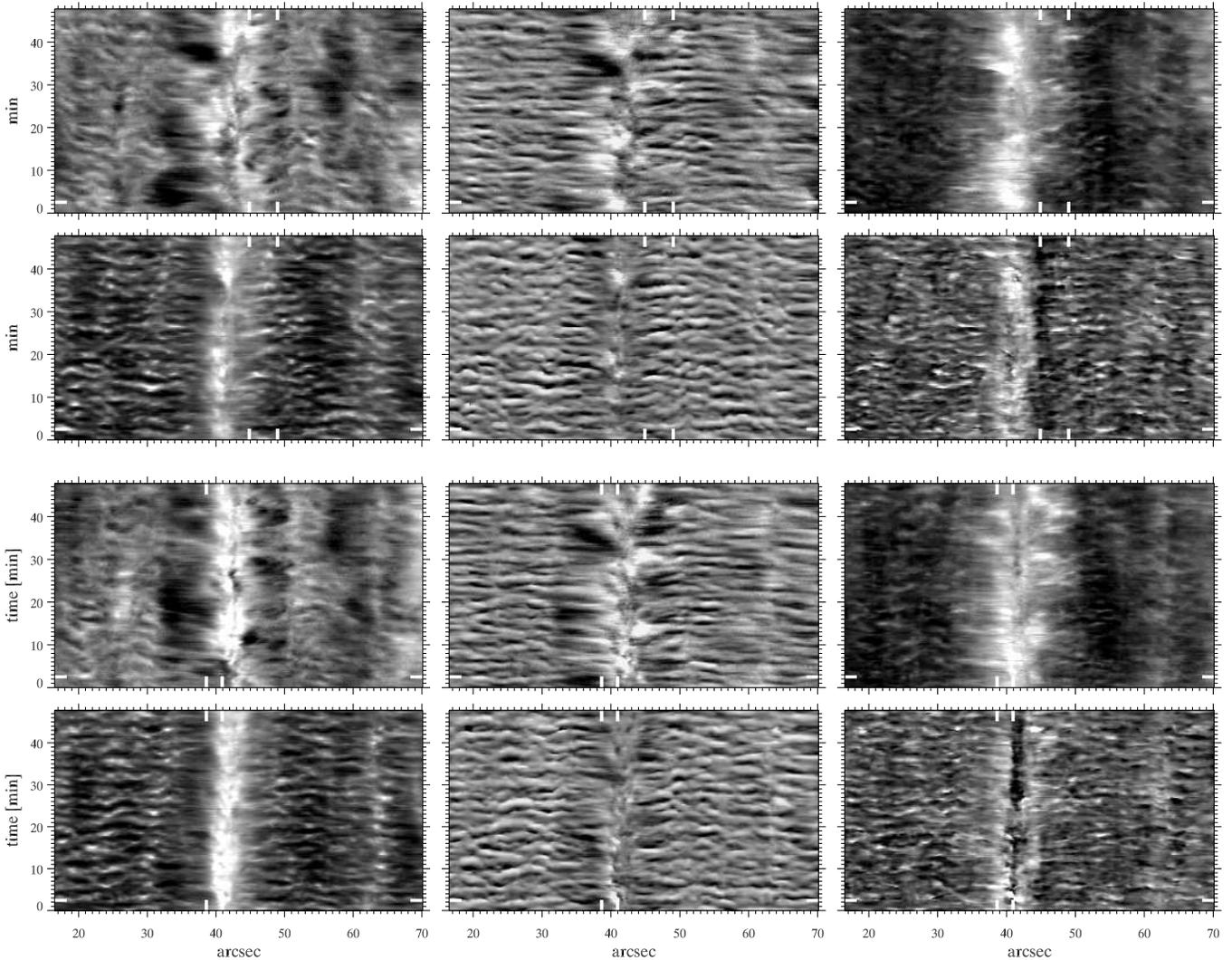}
  \caption[]{
  Space-time slices showing temporal behavior along the two cuts
  in the $x$-direction that are defined in Fig.~\ref{fig:meanimages} by white
  horizontal markers along the $y$-axes.  The upper half of
  this figure corresponds to the higher cut.  The six panels per cut
  show the same quantities as in Figs.~\ref{fig:instimages} and \ref{fig:meanimages}.  Each
  panel is bytescaled separately, the Dopplershifts symmetrically with
  respect to zero.  The white markers along the spatial axes specify
  four pixels, two per slice, for which the spectral profile evolution
  is shown in Fig.~\ref{fig:spectime}.  The white markers near the
  bottom of the time axes specify the moment of observation of the
  profiles in Fig.~\ref{fig:width-measurement} and the parameter
  displays in Fig.~\ref{fig:instimages}.
}
\label{fig:slices}
\end{figure*}
%===========================================================================

%===========================================================================
%% Fig.~\ref{fig:scatter}
%===========================================================================
\begin{figure*}
  \centering 
  \includegraphics[width=88mm]{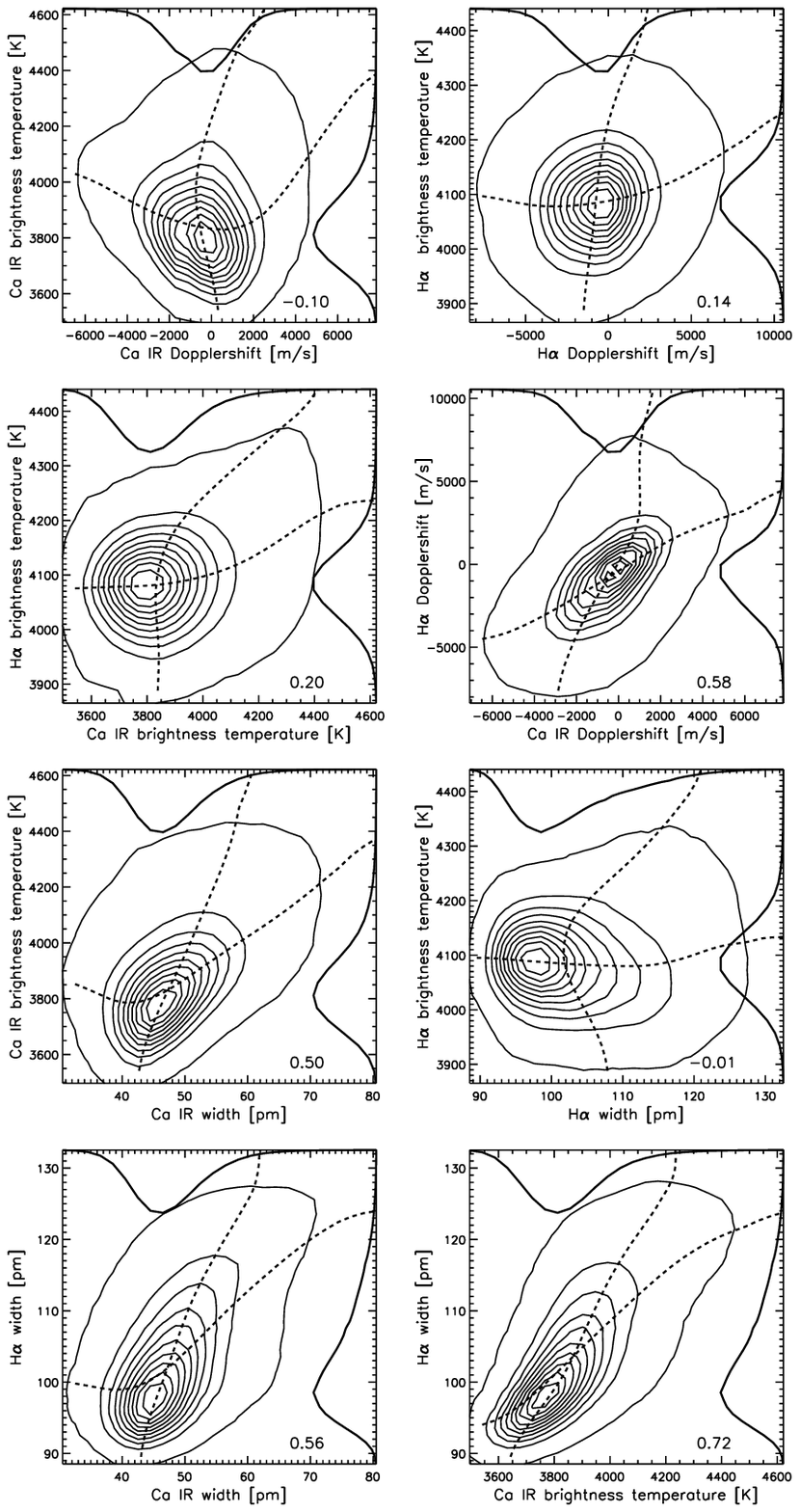}
  \includegraphics[width=88mm]{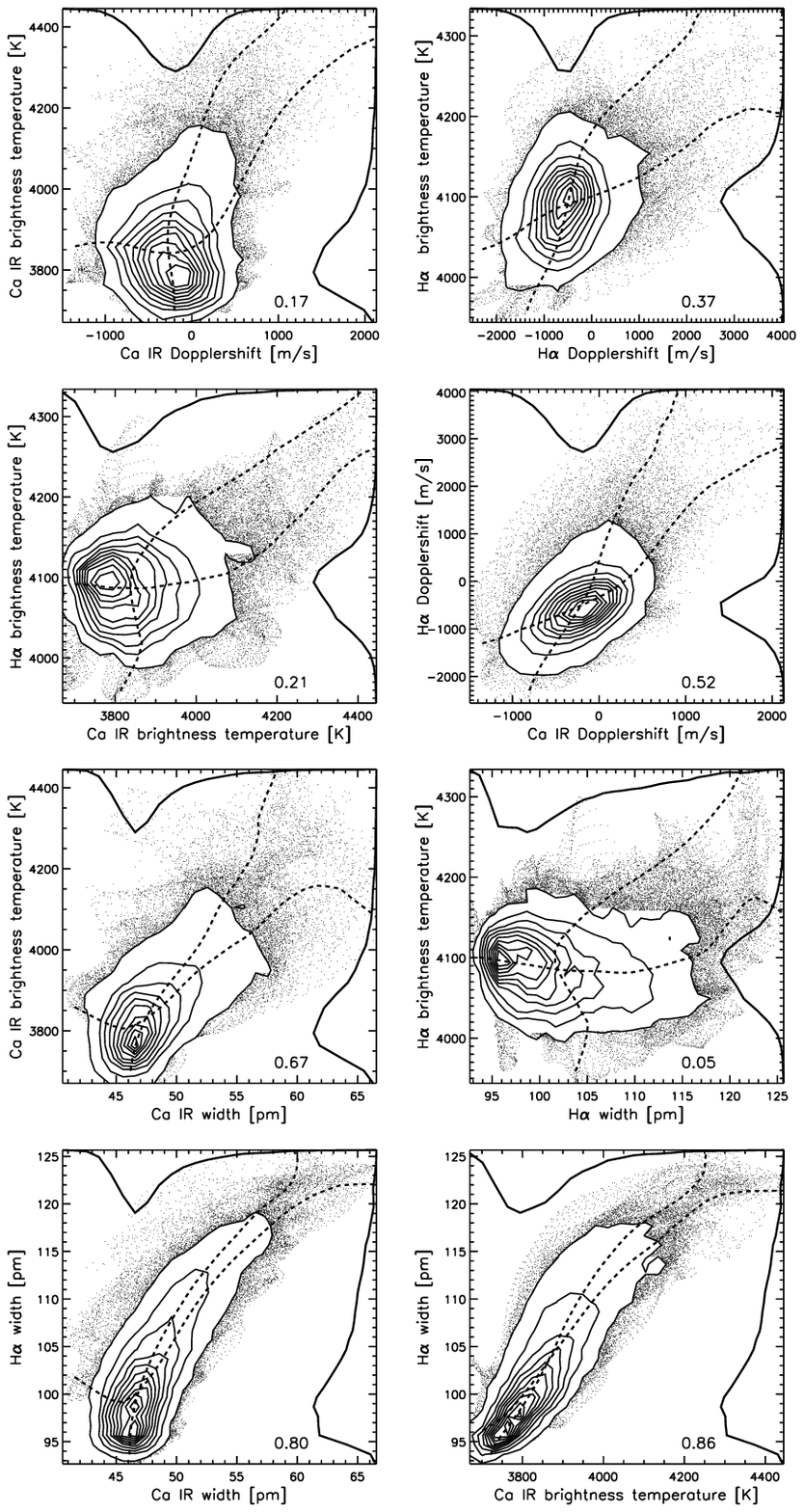}
  \caption[]{
 Scatter-plots  between measured quantities for all pixels in the
  full field of view.  The first two columns are instantaneous
  pixel-by-pixel comparisons between the quantities specified along
  the axes, for the 66 profile scans with the best seeing ($10^7$
  samples).  The third and fourth columns are similar comparisons for
  the same quantities but temporally averaged over the whole sequence,
  excluding the third of the scans with the worst seeing
  ($1.6\times10^5$ samples).  The first two columns correspond to
  Fig.~\ref{fig:instimages}, the other two to
  Fig.~\ref{fig:meanimages} but sampling the full field.  {\em Plot
  format\/}: to avoid plot saturation, single pixel-by-pixel sample
  plotting is replaced by plotting sample density contours, except
  that the extreme individual pixel-by-pixel comparisons beyond the
  outer contour are shown in columns 3 and 4.  Extended arcs in these
  outer scatter clouds come from isolated, non-characteristic
  features.  The solid curves along the top and right sides show the
  occurrence distribution per quantity. The dashed curves specify the
  first moments along cuts in each axis direction.  They align at
  perfect correlation (tilt to the upper right) or anticorrelation
  (tilt to the upper left) and they are perpendicular along the axis
  directions in the absence of any correlation.  The numbers near the
  lower-right corners specify the overall Pearson correlation
  coefficient.  The intensities of the profile minima are
  quantified as brightness temperatures.  Positive Dopplershift
  implies downdraft.  The scales differ between the instantaneous and
  sequence-averaged comparisons, especially for Dopplershift.  The
  contours in the latter are noisier due to the smaller statistics.
  {\em Top row\/}: the intensity of the profile minimum against
  its Dopplershift for \CaIR\ and \Halpha.  {\em Second row\/}:
  intensity--intensity and Dopplershift--Dopplershift comparisons.
  {\em Third row\/}: minimum intensity against core width for
  \CaIR\ and \Halpha.  {\em Bottom row\/}: \Halpha\ core width against
  \CaIR\ core width and against \CaIR\ minimum intensity.
}
\label{fig:scatter}
\end{figure*}
%===========================================================================
%RR merged into one fig because they must be seen together

%%%%%%%%%%%%%%%%%%%%%%%%%%%%%%%%%%%%%%%%%%%%%%%%%%%%%%%%%%%%%%%%%%%%%%%%%%%%
\section{Parameter displays and comparisons}  \label{sec:displays}
%%%%%%%%%%%%%%%%%%%%%%%%%%%%%%%%%%%%%%%%%%%%%%%%%%%%%%%%%%%%%%%%%%%%%%%%%%%%

Figure~\ref{fig:instimages} shows our measurements from a single profile
scan of the two lines, taken at 17:52:34~UT 
%GC I did 10*15.4 s + 17:50. OK?
during one of the best-seeing moments.  For each line the measurements
are as specified in the previous section: the intensity of the line
profile minimum expressed as brightness temperature, its Dopplershift,
and the line core width defined as in Fig.~\ref{fig:width-measurement}.  We
describe them column by column, comparing each panel with pertinent
other ones.  These comparisons are quantified in the form of
pixel-by-pixel correlation diagrams in the lefthand part of
Fig.~\ref{fig:scatter}, which we describe in parallel.  
The diagrams are
constructed from 66 spectral sets taken during very good seeing to
obtain the best statistics.  The caption explains the format, while the
axes provide the numerical ranges of the greyscale variations in
Fig.~\ref{fig:instimages}.  The innermost contours of the diagrams in Fig.~\ref{fig:scatter} represent the maxima of the overall distribution, while 
the curves on the top and right of each figure show the distribution
of the relative quantity on each axis.

The two images of the profile-minimum intensity at left in Fig.~\ref{fig:instimages} differ
dramatically in their portrayal of the solar atmosphere.  The
corresponding scatter diagram (first panel of the second row in
Fig.~\ref{fig:scatter}) indeed shows a total lack of brightness
correlation between the two, except for the very brightest pixels.
The \Halpha\ image shows masses of bright and dark chromospheric
fibrils, sometimes with marked brightening towards one end.
%RK I maintain this - not cospatial with fluxtubes!
Comparison with the magnetograms in Fig.~\ref{fig:GONG-evolution}
shows that these bright endings 
%% are rooted in 
correspond to
the stronger magnetic concentrations, 
but the magnetic network pattern cannot be recognized
directly from the \Halpha\ image.  That is much more easily done in the
\CaIR\ intensity image, in which the extended diffuse bright patches
constituting the chromospheric network correspond reasonably well to
the unsigned magnetogram distribution.  The most active patch is
located at the field edge at about 2 o'clock; it contributes most of
the slight bright-bright correlation in Fig.~\ref{fig:scatter} with
brightness temperature $T_\rmb \approx 4300$~K in both lines.  The
dark internetwork areas contain small-scale brightenings that 
%% evolve very rapidly 
last only a minute or so
and mark acoustic shocks 
%% that we discuss further below 
(see Movie 1 in
\cite{2008A&A...480..515C}; % Cauzzi etal IBIS I
\cite{2009A&A...494..269V}). % Vecchio, Cauzzi & Reardon 2009

We now turn to the Dopplershift maps in the center panels.  The
\Halpha\ Dopplermap has large similarity to the \Halpha\ 
profile-minimum map in its
overall spatial structure.  It shows the same masses of fibrils,
roughly with the same bundle patterns and orientations.  However, the
two are dissimilar in detail: the Dopplermap fibrils tend to be
shorter and they are dark or bright (moving upward or downward)
without obvious relation to the contrast in the profile-minimum image, except for the
brightest and the darkest features whose co-spatiality indicates
exceptional downdrafts in fibril ends near the network.  The lack of
pixel-by-pixel similarity is quantified in the second panel of
Fig.~\ref{fig:scatter} which indeed shows only very slight
correlation.  The corresponding \CaIR\ diagram in the first panel of
Fig.~\ref{fig:scatter} shows slightly higher coherence but with
negative correlation: the intensity tends to be higher at locations
with larger blueshift (dark in the Dopplermap), except for the
brightest pixels which therefore diminish the negative correlation in
the overall Pearson coefficient.  
We show below that the
countercorrelation comes from the internetwork shocks.

Much better coherence is shown in the second panel of the second
row of Fig.~\ref{fig:scatter} which plots the per-pixel correlation
between the two Dopplershift measures.  One would not expect
such high correlation comparing the center-column images in
Fig.~\ref{fig:instimages} since they appear rather dissimilar, but
closer inspection shows that indeed the \CaIR\ Dopplermap has similar
fibrils around network as the \Halpha\ Dopplermap, although less
extended.  In the internetwork the former shows many roundish patches
that must correlate fairly well with the more fibrilar-shaped
Dopplershift features in \Halpha\ to produce the good overall
correlation in the scatter plot.  This correlation may be produced by
acoustic shocks seen simultaneously in both lines, or if
shocks from below buffet \Halpha\ fibrils as suggested by
   \citet{2008SoPh..251..533R} % Rutten++ tomo7 
or excite local turbulence as suggested by
  \citet{2008ApJ...683L.207R}.  % Kevin + Cosenza turbulence tail
The two Dopplermaps so possess larger pixel-by-pixel correlation than
cursory inspection suggests, just the opposite of the \Halpha\
brightness--Dopplermap comparison.

The remaining two panels in Fig.~\ref{fig:instimages} and the
remaining four panels in the lefthand part of Fig.~\ref{fig:scatter}
concern the core width measures of the two lines.  It is most striking
that the \Halpha\ width pattern has much larger similarity to the
\CaIR\ minimum intensity than to anything else; this is the key
point of this paper.  They both show the network as bright patches on
a dark background, with longer fibrilar extensions in \Halpha\ width
than in \CaIR\ brightness.  The internetwork grains stand out more
clearly in \CaIR, but there is very good similarity in the underlying
internetwork patterns.  This large apparent correspondence is
confirmed on the pixel-by-pixel level by the second scatter diagram in
the bottom row of Fig.~\ref{fig:scatter} which shows both
bright-bright and dark-dark correlation with high significance.  There
is also significant correlation between \CaIR\ brightness and width
and between \Halpha\ width and \CaIR\ width in the first column of
Fig.~\ref{fig:scatter}.  There is no correlation at all between
\Halpha\ intensity and width, in accordance with the absence of
correlation between the two brightness images.

Some peculiar patches appear at times in the \CaIR\ parameter images. A clear
example is given by the feature near $(32\arcsec,73\arcsec)$ in Fig.~\ref{fig:instimages}, 
which appears very bright  in the minimum intensity and in the line-width panels, while being very dark (upflow) with a surprising bright center (downflow) in the Dopplershift map. These features often 
represent particularly strong
internetwork shocks with a strongly asymmetric \CaIR\ line core profile
for which our simple measurements are inadequate.  However, such occurrences
are rare (much less than one percent), and do not affect the statistical correlations in 
Fig.~\ref{fig:scatter}.

Figure~\ref{fig:meanimages} shows temporal averages over the sequence
duration of the same quantities as in Fig.~\ref{fig:instimages}.
Moments of less good seeing were excluded by setting a quality
threshold at the mean rms variation of the \Halpha\ profile-minimum
intensity images, discarding nearly one-third of the samples,  spread fairly
evenly in time.  The righthand half of Fig.~\ref{fig:scatter} contains
the corresponding pixel-by-pixel scatter plots for the whole IBIS
field of view, while only the central part of the field is shown in
Fig.~\ref{fig:meanimages}. This is done to enhance the spatial scale in
Fig.~\ref{fig:slices}, which shows the temporal evolution for the
pixels along the two fixed-$y$ cuts indicated by the horizontal white markers in
Fig.~\ref{fig:meanimages}.  They are selected to pass through the two
distinct black patches in the last panel of Fig.~\ref{fig:meanimages}
and are 1.5~arcsec apart.  We describe these figures together.

Figure~\ref{fig:slices} shows that periodic 
phenomena abound.  All time slices show ubiquitous oscillatory
modulation, not only in the internetwork areas but also for the
fibrils that jut out from the small area of active network sampled in
the middle.  The oscillatory signals are most obvious in the
Dopplershift patterns (center column) which are remarkably similar for
the two lines.  They are least clear but still visible in \Halpha\
intensity (first panel).  They average out when taking a
temporal mean, so that the spatial patterns defined by longer-lived
structures and motions gain visibility in Fig.~\ref{fig:meanimages} in
comparison with Fig.~\ref{fig:instimages}.

In all measures the internetwork is full of three-minute oscillations
with shock signatures, clearest in the \CaIR\ intensity panels
in the form of bright grains 
(\cite{2009A&A...494..269V}). % Vecchio++ shocks 8542
These are comparable to the \CaII\ \HtwoV\ and \KtwoV\ grains reviewed
by
   \citet{1991SoPh..134...15R} % Rutten+Uitenbroek review
and explained by
  \citet{1997ApJ...481..500C} %C Carlsson+Stein, H2v grain simulation
as steepening and vertically interacting shock waves that travel up
until wave conversion occurs at the magnetic canopy
  (\cite{2003ApJ...599..626B}). % Bogdan+Oslo waves+canopies
The \Halpha\ time slices show that \Halpha\ partakes in this
three-minute modulation, even in its intensity although that has no
resemblance nor correlation to \CaIR\ in the instantaneous images and
scatter diagrams.  Further study of this modulation calls for Fourier
analysis including phase-difference evaluation with
network/internetwork separation following 
  \citet{2001A&A...379.1052K}, % Krijger++ TRACE1
which we postpone to a future paper.
  
The network patch at the center of the time slices shows oscillatory
behavior of its own, most clearly in the Dopplershift panels.  Some
fibrils that jut out from it, especially towards the left in \Halpha,
have alternating flows and may represent dynamic fibrils
as described by
  \citet{2006ApJ...647L..73H} % Hansteen++ dynamic fibrils
and
  \citet{2007ApJ...655..624D}. % DePontieu++ dynamic fibrils  
The central part displays slower undulations as described 
by
  \citet{1993ApJ...414..345L}. %T Lites+Rutten+Kalkofen, network dynamics
The fibril-dominated regions surrounding the network show a distinctly 
different behavior than either the nework or internetwork regions, as described 
by \citet{2009A&A...494..269V}. % Vecchio et al. 2009 shocks

The temporal averaging in Fig.~\ref{fig:meanimages} and the righthand
part of Fig.~\ref{fig:scatter} reduces the spread due to these
oscillations and other short-lived phenomena, and so serves to enhance
longer-term structuring and relationships
%GC 011208 add:
(note also the reduced ranges of the axes in the right-hand side of Fig.~\ref{fig:scatter}).
The averaging over the
oscillatory components indeed reduces the blue-bright correlation 
%GC 011208 add:
between \CaIR\ brightness and Dopplershift
in the first panel of Fig.~\ref{fig:scatter},
% same
introduced by internetwork acoustic shocks. 
In the temporal averaging of the \Halpha\ data, 
the fibril contribution gains dominance and improves the correlation
with intensity, corresponding to larger resemblance between the first
two \Halpha\ panels in Fig.~\ref{fig:meanimages} than in
Fig.~\ref{fig:instimages}.

The scatter plots in the second and third rows of
Fig.~\ref{fig:scatter} do not change much in the temporal averaging.
The Dopplershifts comparison in the second row looses the internetwork
oscillations but maintains fairly good correlation, indicating
a correspondence in the fibril flows also seen in the center panels of
Fig.~\ref{fig:meanimages}.  In the third row there is slight
improvement in the high-high correlation between
\CaIR\ brightness and width.

The most striking change that the temporal averaging brings is
increased similarity between the panels relative to the
\Halpha\ width, \CaIR\ profile-minimum intensity and \CaIR\ width of
Fig.~\ref{fig:meanimages} in comparison with
Fig.~\ref{fig:instimages}.
In particular, the time-averaged \Halpha\
width (third panel in upper row) shows a pattern of smooth, bright network aureoles
on a dark internetwork background that appears very similar to the
\CaIR\ intensity (first panel in lower row).  The bright network patches seem
smaller in the latter, but this results from the non-linearity
displayed in the last panel of Fig.~\ref{fig:scatter}, which quantifies
this comparison.  It shows high correlation along the whole
scatter-sample mountain, but the turnover of the upward tail implies
that the brightest network patches appear larger in \Halpha\ width.
The whole mountain ridge is appreciably narrower in the time-averaged
correlation (last panel) than in the instantaneous one (bottom second
column).
%RR note that x-scale is also more extended
Thus, both network (contributing the upper part) and internetwork
(lower part) gain pattern similarity from temporal averaging in this
comparison.  This tight relationship suggests a common physical origin,
which we identify as temperature below.  The width-width correlation
(bottom panel of the third column in Fig.~\ref{fig:scatter}) is
just as good as the \Halpha-width vs.\ CaII\ intensity correlation
for large core widths but it vanishes for small core widths, 
indicating that temperature dominates the \CaIR\ width only when high
and that temporal averaging over the internetwork shocks weakens their
temperature signature, perhaps through their short duration with
respect to their periodicity.

%===========================================================================
%% Fig.~\ref{fig:spectime}
%===========================================================================
\begin{figure}
%%  \sidecaption
  \centering \includegraphics[width=88mm]{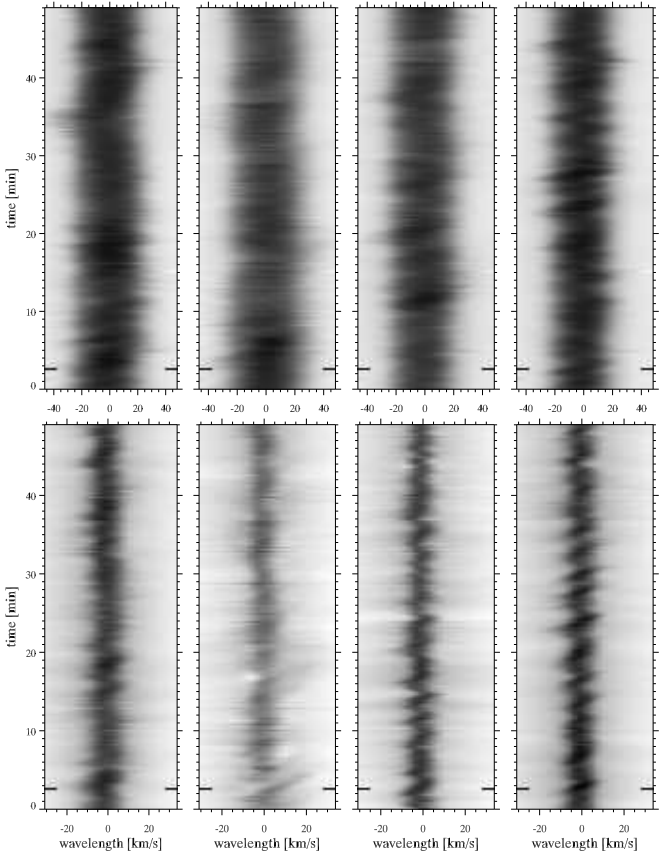}
  \caption[]{
  Spectrum-versus-time plots showing the spectral behavior of \Halpha\
  (upper row) and \CaIR\ (lower row) for the four selected pixels
  specified by markers in Figs.~\ref{fig:meanimages} and
  \ref{fig:slices}.  The wavelengths along the abscissae are expressed
  as Dopplershift from the nominal line center.  The logarithm of the
  spectral brightness is shown with the same greyscale for all panels
  along each row; it is logarithmic in order to detail line-core
  variations, especially for \Halpha\ which is too uniformly dark
  otherwise.  The first and second columns are for pixels along the
  lower cut in Figs.~\ref{fig:meanimages} and \ref{fig:slices}.  The
  first pixel samples normal network, the second the network feature
  with low \CaIR\ core width.  The third and fourth columns are for
  pixels along the upper cut and sample the near-network feature with
  low \CaIR\ core width and internetwork further away.  The black
  markers near the bottom of each panel specify the time of
  observation selected in Figs.~\ref{fig:width-measurement} and
  \ref{fig:instimages}.  The second pixel happened to lie in a dynamic
  fibril at that moment, causing large \CaIR\ redshift and blue-wing
  brightening also seen in Fig.~\ref{fig:width-measurement}.
}
\label{fig:spectime}
\end{figure}
%===========================================================================

Note that the \CaIR\ width image in Fig.~\ref{fig:meanimages} shows
dark centers in some of the bright network patches.  The yet brighter
network patches near 2~o'clock and 6~o'clock at the edge of the IBIS
field of view, not included in Fig.~\ref{fig:meanimages}, have similar
dark cores.  The lower slice in Fig.~\ref{fig:slices} samples the
darkest network core in Fig.~\ref{fig:meanimages}.  The upper
slice cuts through a comparably dark feature in \CaIR\ width that lies
adjacent to the network patch.  However, comparison of the two dark \CaIR\ width
patches in the time slices in Fig.~\ref{fig:slices} (the two last
panels of each half) suggests that they have intrinsically different
nature.  The network core patch in the lower panel lies at the center
of the fibril rosette and is partially visible also in \Halpha\ width.
The dark patch adjacent to the network in the upper panel instead 
seems to be regularly oscillating internetwork,
only darker overall.
  
Figure~\ref{fig:spectime} adds further information on these features by comparing the
spectral behavior of the two lines with time, respectively for pixels
within these two dark patches (center columns) and for comparable
pixels elsewhere.  The two left-most columns  sample network
pixels, the two right-most ones sample internetwork.  
The first thing to note about the \Halpha\ line core is its 
much larger width 
with respect to \CaIR, and its ubiquitous presence (note also that the spectral scale of 
Fig.~\ref{fig:spectime} is different for the two lines, covering about 0.2 nm around \Halpha\ and 0.18 nm around \CaIR).
This basal core width is widest in the first
panel, while the superimposed
Doppler excursions become a major component only in the internetwork
(rightmost columns).   The \CaIR\ core shows these Doppler excursions as well, 
but more visibly due to its reduced width.
Regular shock trains occur most clearly in the
internetwork panels for the \CaIR\ line and are also 
recognizable in the corresponding \Halpha\
panels.  In both dark \CaIR-width patches (second and third
columns) the \CaIR\ core indeed appears generally narrower, but also
less dark than for the comparison pixels in the first and last
columns.  Profile inspection along the second column shows that in these pixels
\CaIR\ often has a very shallow chromospheric core for which our width
algorithm produces small values.

Finally, we note that the time of observation selected for
Fig.~\ref{fig:width-measurement} happened
to sample a feature indicative of dynamic fibrils in the second panel of Fig.~\ref{fig:spectime},
\ie\ displaying large blue-to-red successive shifting in \CaIR.  It has
correspondingly large redshift and blue-wing brightening in
Fig.~\ref{fig:width-measurement}.  There is no clear signature of this
feature in \Halpha.

%===========================================================================
%% Fig.~\ref{fig:sourcefunctions}
%===========================================================================
\begin{figure}
  \centering
  \includegraphics[width=80mm]{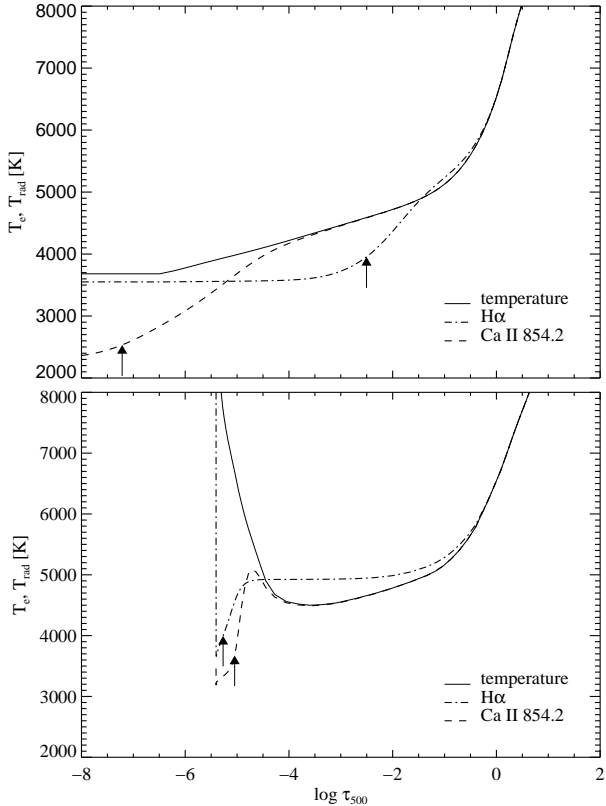}
  \caption[]{
  Demonstration of difference in temperature sensitivity between
  \CaIR\ and \Halpha.  Each panel shows the temperature stratification
  of a standard solar model atmosphere and the resulting total source
  functions $S_\lambda$ at the nominal line-center wavelengths for
  \CaIR\ and \Halpha, as function of the continuum optical depth at
  $\lambda \is 500$~nm and with the source functions expressed as
  formal temperatures through Planck function inversion.
%RR The mean intensity 
%RR $J_\lambda$ is plotted as well for both lines, with open symbols. 
  The arrows mark $\tau \is 1$ locations.
  {\em Upper panel\/}: radiative-equilibrium model KURUCZ from 
  Kurucz (1979, 1992a, 1992b).
    \nocite{1979ApJS...40....1K} % Kurucz models
    \nocite{1992RMxAA..23..181K} % Kurucz solar update
    \nocite{1992RMxAA..23..187K} % Kurucz solar update
  It was extended outward assuming constant temperature in order to
  reach the optically thin regime in \CaIR.  
  {\em Lower panel\/}: empirical continuum-fitting model FALC of
   \citet{1993ApJ...406..319F}. % FAL including FALC model
  Its very steep transition region lies beyond the top of the panel
  but causes the near-vertical source function increases at left.
}
\label{fig:sourcefunctions}
\end{figure}
%===========================================================================

%===========================================================================
%% Fig.~\ref{fig:wwcurve}
%===========================================================================
\begin{figure}
  \centering
  \includegraphics[width=60mm]{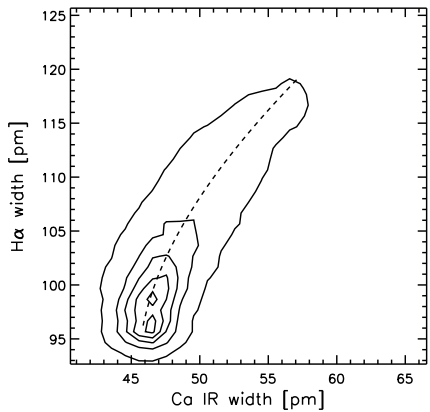}
  \caption[]{
  Width-width correlation between \Halpha\ and \CaIR.  The contours
  reproduce every second one in the next-to-last panel of
  Fig.~\ref{fig:scatter}.  The dashed curve represents
  Eq.~(\ref{eq:width}) with best-fit parameters.
}
\label{fig:wwcurve}
\end{figure}
%===========================================================================

%%%%%%%%%%%%%%%%%%%%%%%%%%%%%%%%%%%%%%%%%%%%%%%%%%%%%%%%%%%%%%%%%%%%%%%%%%%%
\section{Discussion}   \label{sec:discussion}
%%%%%%%%%%%%%%%%%%%%%%%%%%%%%%%%%%%%%%%%%%%%%%%%%%%%%%%%%%%%%%%%%%%%%%%%%%%%

%%%%%%%%%%%%%%%%%%%%%%%%%%%%%%%%%%%%%%%%%%%%%%%%%%%%%%%%%%%%%%%%%%%%%%%%%%%%
\subsection{Temperature sensitivity}                 \label{sec:sensitivity}
%%%%%%%%%%%%%%%%%%%%%%%%%%%%%%%%%%%%%%%%%%%%%%%%%%%%%%%%%%%%%%%%%%%%%%%%%%%%
The \CaIR\ and \Halpha\ lines differ in two significant ways in their
temperature sensitivity.  First, \CaIR\ is much closer to LTE both
with regards to opacity and to source function behavior.  The latter
difference is demonstrated in Fig.~\ref{fig:sourcefunctions} which
represents a didactic step beyond the LTE comparison in Fig.~6 of
  \citet{2006A&A...449.1209L}, %C Leenaarts BPs blue wing Ha
by comparing standard NLTE formation for the two lines.
It shows dramatic difference between the two lines in their
sensitivity to the temperature in the outer atmosphere.  In each case
the lines are computed in NLTE with the RH code of
  \citet{2001ApJ...557..389U}, % Uitenbroek 2001
using a sufficiently complete model atom and accounting for coherent
scattering in Ly\,$\alpha$, Ly\,$\beta$ and \HK\ while assuming complete
redistribution for all other lines.
In the upper panel, for the KURUCZ radiative-equilibrium model \Halpha\
is a deep-photosphere line of which the source function decouples from
the Planck function as deep as $\log \tau_{500} \!\approx\! 0$, together with
the Balmer continuum.  The latter, \Halpha, and \Lyalpha\  combine
together in setting the pertinent hydrogen populations through
intricate interactions involving Balmer continuum pumping, \Halpha\
photon losses, Rydberg ladder flow, and detailed balancing in
\Lyalpha\
  (see \cite{1994IAUS..154..309R}). %T Tucson IAU IR, formation Rydberg lines
The \Halpha\ source function $S_\lambda$ closely follows the mean
intensity $J_\lambda$ (not shown) as if it were an ordinary two-level
scattering line.  Hence, its temperature sensitivity responds mostly
to deep-photosphere perturbations.  In contrast, the \CaIR\ line has
much larger opacity, so that its source function follows the
temperature out to $\log \tau_{500} \approx -4$ before the onset of
its scattering drop with $S_\lambda \approx J_\lambda$ which makes the
line very dark.

In the lower panel the photospheric part of the FALC temperature
stratification is closely the same as the KURUCZ one
%RR Bartolomeo's plot in MnI FALC-KURUZ = deep parts against log m
and so the formation of the lines there remains similar, with
thermalization out to $\log \tau_{500} \!\approx\! 0$ for
\Halpha\ and $\log \tau_{500} \!\approx\! -4$ for \CaIR.
The addition of the FALC chromosphere results in a source function
bulge for \CaIR\ that is familiar from the older explanations of
\CaII\ \HK\ reversals before partial redistribution and shock dynamics
were included
  (\eg\ Fig.~13 of \cite{1970PASP...82..169L}). % Avrett+Linsky HK
For \Halpha, the chromosphere plus transition region act as the
addition of a barely opaque, hot, radially inhomogeneous slab well
above the photosphere.  It raises the scattering source function
across the upper photosphere but without giving it sensitivity to the
local temperature since there is virtually no local
\Halpha\ opacity there.

%RH Halpha opacity also more out of LTE than CaIR - but wrong anyhow in SE

The $\tau \is 1$ arrows in Fig.~\ref{fig:sourcefunctions} indicate
that the predicted line-center brightness temperatures $T_\rmb \approx
T_{\rm rad}(\tau_\lambda\is1)$ happen to be nearly equal for \Halpha\
at $T_\rmb \!\approx\! 3900$~K for both models, but in very different
manner.  The photosphere-only prediction for \CaIR\ line-center
brightness ($T_\rmb \!\approx\! 2500$~K) is instead much lower 
than with a chromosphere ($T_\rmb \!\approx\! 3600$~K).  The
observed mean values from Fig.~\ref{fig:scatter} are $T_\rmb \!\approx\!
4100$~K for \Halpha\ and $T_\rmb \approx 3800$~K for \CaIR.  
Given the
very different structure of the source function of the two lines, the
closeness of their two brightness temperatures in the model with a chromosphere
is more likely a coincidence imposed by the atmospheric model than
proof that the lines supply temperature information from the same
height in the atmosphere.  The \Halpha\ line-minimum intensity
is much more sensitive to deep-photosphere temperature perturbations,
whereas the \CaIR\ line minimum has more sensitivity to
chromospheric temperature perturbations.  The first columns of
Figs.~\ref{fig:instimages} and
\ref{fig:meanimages} confirm this difference dramatically by
displaying very dissimilar chromospheric intensity scenes, not only in
the instantaneous comparison (Fig.~\ref{fig:instimages}) but also in
the time-averaged one (Fig.~\ref{fig:meanimages}).  In contrast, the
closeness of the formation heights in the chromospheric model suggests
that the Dopplershifts and Doppler broadening, which are encoded
locally at the last photon scattering independent of the source
function nature, should be much more similar.  This may explain
the larger similarity between the two lines in the corresponding
panels of Figs.~\ref{fig:instimages} -- \ref{fig:scatter}.
However, the large difference in \Halpha\ formation height
between the KURUCZ and FALC models on the one hand and the 
larger apparent extent of fibrils in \Halpha\ than in \CaIR\
in  Figs.~\ref{fig:instimages} and \ref{fig:meanimages}
on the other hand imply that there is no guarantee
of Doppler-encoding similarity between the two lines.

The second major difference between the two lines is that they differ
by a factor of 40 in atomic mass.  This affects the thermal Doppler
broadening and therefore the balance between thermal and nonthermal
Doppler broadening of the line profile.  The latter was traditionally described by the
ad-hoc fudge parameters microturbulence and macroturbulence, with the
former entering the Dopplerwidth and the latter applied as Gaussian
smearing of the computed emergent line profile.  In recent times the
usage of these parameters has been replaced by structure-resolving
line synthesis from numerical simulations, but so far only for granular
convection and magnetoconvection in the low photosphere.
Given the low atomic mass, \Halpha\ is likely to have thermal broadening
dominate over systematic motions, whereas the reverse holds for \CaIR.
The spectrum-versus-time displays in Fig.~\ref{fig:spectime} confirm
this difference qualitatively.  Hence, we attribute the large,
generally-present core width of \Halpha\ to thermal broadening, and
therefore the bright patches in the third panel of
Fig.~\ref{fig:meanimages} to high temperatures which set the core
width locally, at the last scattering.  The good correspondence of the \Halpha\ line width with
the patches of high \CaIR\ profile-minimum intensity in the fourth panel then results from the appreciable
\CaIR\ source function sensitivity to temperature.

Figure~\ref{fig:wwcurve} quantifies this conclusion in the form of
a simple model which reproduces the observed correlation in the next-to-last
panel of  Fig.~\ref{fig:scatter}.  The dotted line in Figure plots the widths of the two lines
against each other that were computed from
\begin{equation}
  \Delta \lambda_\rmW 
     \equiv (\lambda/c) \sqrt{v_{\rm therm}^2 + v_{\rm micro}^2 
                                 + v_{\rm intrins}^2} 
  \label{eq:width}
\end{equation}
for temperature $T$ increasing from $T_{\rm min} = 5\,000$~K to
$T_{\rm max}=60\,000$~K, a thermal contribution $v_{\rm therm} =
\sqrt{2kT/m}$ where $k$ is the Boltzmann constant and $m$ the atomic
mass, a microturbulent contribution $v_{\rm micro} = 1 +
10~(T\!-\!T_{\rm min})/(T_{\rm max}\!-\!T_{\rm min})$~\kms, and an
intrinsic contribution $ v_{\rm intrins}$ set to 43~\kms\ for \Halpha\
and 16~\kms\ for \CaIR.  The latter two contributions were adapted to fit the
observed correlation.  The temperature-dependent microturbulence
increases from 1~\kms\ to 11~\kms, in rough agreement with older
estimations from ultraviolet line broadening
  (\cite{1976pmas.conf..291C}; % Canfield+Beckers Nice
  \cite{1978ApJ...220..314T}; % Tripp++ Si lines 
  \cf\ Fig.~11 of
  \cite{1981ApJS...45..635V}). % VALIII 
The intrinsic widths represent base contributions making up for the
neglect of radiative transfer.  Eq.~(\ref{eq:width}) may be seen as
describing line broadening in a Schuster-Schwarzschild slab or cloud
with the irradiative profile and internal line transfer mimicked by
these intrinsic widths.  It is only a simple ad-hoc fit, but it
demonstrates that the observed mean width-width correlation in
Fig.~\ref{fig:scatter} may be attributable to temperature.  
The value
$T_{\rm max} \is 60\,000$~K seems very high in comparison with
standard one-dimensional models in which both hydrogen and
once-ionized calcium are fully ionized at such temperatures, but
observations as in Fig.~13 of
  \citet{2007ASPC..368...27R} % Rutten Coimbra
suggest the presence of \Halpha\ even at very high temperature. 

%RR 171 and Halpha are both bright at fibril foots in ARs = straws I think

Note that such core width behavior cannot be reproduced by
specification of a deeper temperature rise as done by
   \citet{1993ApJ...406..319F}. % FAL including FALC model
Their A, C, \ldots, F and P models were designed to represent
progressively hotter chromospheres in quiet cell interior, average
quiet sun, network, and plage by placing the chromospheric temperature
rise at increasingly larger column mass.  However, when 
measuring with our method the line widths
from profiles synthesized in NLTE (as in Fig.~\ref{fig:sourcefunctions}),  
this sequence produces smaller
core width with increasing activity instead of
the actual core-width increase we observe.

%%%%%%%%%%%%%%%%%%%%%%%%%%%%%%%%%%%%%%%%%%%%%%%%%%%%%%%%%%%%%%%%%%%%%%%%%%%%
\subsection{Chromospheric heating} \label{sec:heating}
%%%%%%%%%%%%%%%%%%%%%%%%%%%%%%%%%%%%%%%%%%%%%%%%%%%%%%%%%%%%%%%%%%%%%%%%%%%%

The demonstrations above make us conclude that the high correlation
between time-averaged \Halpha\ core width and \CaIR\ line-minimum
intensity represents common sensitivity to temperature.  Hence, the
network patches display excess heat.  The general, basal presence of a
wide \Halpha\ core in the first columns of Fig.~\ref{fig:spectime}, as
well as the large \Halpha\ widths ubiquitously present in the network in
Fig.~\ref{fig:slices} indicate that this
%RG latex nono!  always Fig.~, not without tilde (too much white)
heating is rather constant in time in the magnetic concentrations.  Temporal
averaging over episodes of strong heating occurring at the base of the
fibrils jutting out from the network (cf.\ the rightmost panels of
%RG latex nono!  always slash after period after small letter
Fig.~\ref{fig:slices}) produces the smooth bright network aureoles in
the third panel of Fig.~\ref{fig:meanimages}.  The dark network patch
of small \CaIR\ width sampled in Figs.~\ref{fig:slices} and
Fig.~\ref{fig:spectime} indicate reduce core-widths related 
to a significant weakening of the line core, 
likely marking maximum heating at the center of
the network patch.

The calculated fit to the width-width correlation
in Fig.~\ref{fig:wwcurve} includes a sizable
microturbulence component which scales with temperature.  Its
necessity suggests the action of unresolved waves. Perhaps these are
the ones affecting or producing the ``straws'' or ``spicules-II''
observed in
\CaIIH, respectively near the limb by
  \citet{2006ASPC..354..276R} % Rutten Steinfest 
and across the limb by
  \citet{2007PASJ...59S.655D}, % BdP++ spicules-II
that were interpreted as Alfv\'en waves by
  \citet{2007Sci...318.1574D}. % BdP++ spicules-II = Alfven
They may also explain unresolved ultraviolet line broadening
  (\cite{2008ApJ...673L.219M}) % McIntosh++ UV line widths = spicules-II
and the ``rapid blueshift events'' at the edge of network
concentrations reported by
\citet{2008ApJ...679L.167L}. %Langangen++ spicules-II 
It is also interesting to note that the higher-temperature regions (i.e. network 
and fibrils) display a stronger turbulence spectrum of  the \CaIR\ velocity than the internetwork \citep{2008ApJ...683L.207R}.

The similarity between \Halpha\ width and \CaIR\ intensity extends to
the oscillation sequences in the internetwork areas in
Fig.~\ref{fig:slices}.  The \Halpha\ width shows internetwork grains
less markedly than \CaIR\ intensity, but it follows the same
modulation.  The bright phases may therefore be interpreted as 
localized temperature increases, but
the overall darkness of the internetwork in Fig.~\ref{fig:meanimages}
implies less total heating than in the network.  The dark, near-network
patch in \CaIR\ width sampled in Figs.~\ref{fig:slices} and
Fig.~\ref{fig:spectime} also seems to have smaller \CaIR\ width from
being brighter than the comparison internetwork pixel, perhaps from
more sub-canopy heating so close to network.

%%%%%%%%%%%%%%%%%%%%%%%%%%%%%%%%%%%%%%%%%%%%%%%%%%%%%%%%%%%%%%%%%%%%%%%%%%%%
\subsection{Future interpretation} \label{sec:cloudmodeling}
%%%%%%%%%%%%%%%%%%%%%%%%%%%%%%%%%%%%%%%%%%%%%%%%%%%%%%%%%%%%%%%%%%%%%%%%%%%%

The next step in profile interpretation is to distill more
intrinsic line formation parameters.  Traditionally, chromospheric
fine structure is described through cloud modeling
 (see review by \cite{2007ASPC..368..217T}). % Tziotziou Coimbra
In its most basic form it delivers the four parameters: optical
thickness, source function (sometimes parabolic), Dopplershift, and
Dopplerwidth for a homogeneous cloud irradiated from below.  Our
finding that in particular \Halpha\ profile-minimum intensity has no
correlation with the actual temperature (as sampled by \Halpha\ width)
nor with the instantaneous \Halpha\ Dopplershift implies that the rich
fibrilar structure displayed by this line in narrow-band
filtergrams is indeed a complex mixture of all four parameters.  A
complication adding uncertainty is the required specification of the
\Halpha\ profile in the cloud irradiation from below and aside.
Figure~\ref{fig:sourcefunctions} suggests that combination of
radiative-equilibrium modeling with back-radiation from the cloud may
serve as first approximation.

Granted cloud-model determinations of these four parameters, the next
issue is how to interpret the cloud source function.  Already in his
original formulation for off-limb spicules,
  Beckers (\eg\ 1964, 1968, 1972)
    \nocite{Beckers1964} % thesis Utrecht
    \nocite{1968SoPh....3..367B} % Beckers spicule review 1
    \nocite{1972ARA&A..10...73B} % Beckers spicule review 2
%%    \nocite{2000eaa..bookE2019B} % Beckers spicule review 3
%
converted the basic empirical cloud parameters into density
and temperature through the tables of
    \citet{1967AuJPh..20...81G} % Giovanelli H + Ca in chrom (basis Beckers)
for NLTE hydrogen ionization.  More recent reformulations of this
approach
  (\eg\
   \cite{1990A&A...230..200A}; %C Alissandrakis++
   \cite{1997A&A...324.1183T}) %C Tsiropoula+Schmieder params
have generally been replaced by large-volume NLTE line profile
computation followed by data ``inversion'' through best-fit parameter
adaptation
  (see \cite{2007ASPC..368..217T} % Tziotziou Coimbra
   for references).
However, 
    \citet{2007A&A...473..625L} %C Leenaarts hion2
have undermined this technique by finding that in chromospheric
structures which suffer repetitive episodic heating, the actual
populations of the atomic levels governing \Halpha\ can be very far
from time-independent statistical equilibrium.  Since shocks seem
ubiquitously present in the chromosphere (center column of
Fig.~\ref{fig:slices}), \Halpha\ cloud-model interpretation needs to
be reformulated including time-dependent hydrogen balancing.  A first
approximation may be to maintain the high degree of ionization and the
corresponding large \Halpha\ population fractions reached at moments
of high temperature.

Obviously, it will be even better to compare data as ours directly to
numerical spectral line synthesis employing ab-initio MHD
simulations of time-dependent chromospheric fine structure, but these
are not yet available (apart from unrealistic ones assuming LTE even
in the chromosphere).
When realistic ones with proper \CaIR\ and \Halpha\ synthesis arise,
reproducing the scatter diagrams in Fig.~\ref{fig:scatter} will provide an
excellent proving ground.

%%%%%%%%%%%%%%%%%%%%%%%%%%%%%%%%%%%%%%%%%%%%%%%%%%%%%%%%%%%%%%%%%%%%%%%%%%%%
\section{Conclusion}  
%%%%%%%%%%%%%%%%%%%%%%%%%%%%%%%%%%%%%%%%%%%%%%%%%%%%%%%%%%%%%%%%%%%%%%%%%%%%

Imaging spectroscopy with IBIS provides diagnostics well beyond
slit-limited spectroscopy or bandpass-limited filter imaging.  The
long-duration simultaneous \CaIR\ and \Halpha\ profile-resolved IBIS
sequences presented here give direct evidence of chromospheric heating
in network.  The clincher measurement is the \Halpha\ core width.

The network heating appears to occur fairly continuously and to affect
also the nearby base of fibrilar features jutting out from it, but it
does not extend as far as, nor shows the fibrilar fine structure of,
the petals of classical rosettes in \Halpha\ filtergrams.  Their
visibility arises from complex combinations of Doppler width,
Dopplershift, optical thickness, and source function variations,
without direct coupling to fibril temperature.  In contrast, the
striking correspondence between \Halpha\ core width and
\CaIR\ line-minimum brightness (Figs.~\ref{fig:meanimages} and
\ref{fig:scatter}) proves that the latter is a better temperature
proxy.  The best, however, is the \Halpha\ core width.

%%%%%%%%%%%%%%%%%%%%%%%%%%%%%%%%%%%%%%%%%%%%%%%%%%%%%%%%%% ACKNOWLEDGEMENTS
\begin{acknowledgements}
  We are much indebted to DST observers Mike Bradford, Joe Elrod and
  Doug Gilliam. IBIS was constructed by INAF/OAA with contributions
  from the University of Florence, the University of Rome, MIUR, and
  MAE, and is operated with support of the National Solar Observatory.
  The NSO is operated by the Association of Universities for Research
  in Astronomy, Inc., under cooperative agreement with the National
  Science Foundation.  R.J.~Rutten, G. Cauzzi, and K.P. Reardon thank the National Solar
  Observatory/Sacramento Peak for their ever-present hospitality, and the Leids
  Kerkhoven-Bosscha Fonds for travel support.  This research made much
  use of NASA's Astrophysics Data System.
\end{acknowledgements}

%%%%%%%%%%%%%%%%%%%%%%%%%%%%%%%%%%%%%%%%%%%%%%%%%%%%%%%%%%%%%%%% REFERENCES
%%\bibliographystyle{aabib} 

%\bibliographystyle{aa}
%\bibliography{aajour,/tmp/rjrfiles.bib,/tmp/adsfiles.bib}

\begin{thebibliography}{51}
\expandafter\ifx\csname natexlab\endcsname\relax\def\natexlab#1{#1}\fi

\bibitem[{{Alissandrakis} {et~al.}(1990){Alissandrakis}, {Tsiropoula}, \&
  {Mein}}]{1990A&A...230..200A}
{Alissandrakis}, C.~E., {Tsiropoula}, G., \& {Mein}, P. 1990, \aap, 230, 200

\bibitem[{Beckers(1964)}]{Beckers1964}
Beckers, J.~M. 1964, A Study of the Fine Structures in the Solar Chromosphere
  (AFCRL Environmental Research Paper No.~49: PhD thesis Utrecht University)

\bibitem[{{Beckers}(1968)}]{1968SoPh....3..367B}
{Beckers}, J.~M. 1968, \solphys, 3, 367

\bibitem[{{Beckers}(1972)}]{1972ARA&A..10...73B}
{Beckers}, J.~M. 1972, \araa, 10, 73

\bibitem[{{Bello Gonz{\'a}lez} \& {Kneer}(2008)}]{2008A&A...480..265B}
{Bello Gonz{\'a}lez}, N. \& {Kneer}, F. 2008, \aap, 480, 265

\bibitem[{{Bogdan} {et~al.}(2003){Bogdan}, {Carlsson}, {Hansteen}, {McMurry},
  {Rosenthal}, {Johnson}, {Petty-Powell}, {Zita}, {Stein}, {McIntosh}, \&
  {Nordlund}}]{2003ApJ...599..626B}
{Bogdan}, T.~J., {Carlsson}, M., {Hansteen}, V.~H., {et~al.} 2003, \apj, 599,
  626

\bibitem[{{Canfield} \& {Beckers}(1976)}]{1976pmas.conf..291C}
{Canfield}, R.~C. \& {Beckers}, J.~M. 1976, in Physique des Mouvements dans les
  Atmospheres, ed. R.~{Cayrel} \& M.~{Steinberg}, 291

\bibitem[{{Carlsson}(2007)}]{2007ASPC..368...49C}
{Carlsson}, M. 2007, in Astronomical Society of the Pacific Conference Series,
  Vol. 368, The Physics of Chromospheric Plasmas, ed. P.~{Heinzel},
  I.~{Dorotovi{\v c}}, \& R.~J. {Rutten}, 49

\bibitem[{{Carlsson} \& {Stein}(1997)}]{1997ApJ...481..500C}
{Carlsson}, M. \& {Stein}, R.~F. 1997, \apj, 481, 500

\bibitem[{{Cauzzi} {et~al.}(2008){Cauzzi}, {Reardon}, {Uitenbroek},
  {Cavallini}, {Falchi}, {Falciani}, {Janssen}, {Rimmele}, {Vecchio}, \&
  {W{\"o}ger}}]{2008A&A...480..515C}
{Cauzzi}, G., {Reardon}, K.~P., {Uitenbroek}, H., {et~al.} 2008, \aap, 480, 515

\bibitem[{{Cavallini}(2006)}]{2006SoPh..236..415C}
{Cavallini}, F. 2006, \solphys, 236, 415

\bibitem[{{De Pontieu} {et~al.}(2007{\natexlab{a}}){De Pontieu}, {Hansteen},
  {Rouppe van der Voort}, {van Noort}, \& {Carlsson}}]{2007ApJ...655..624D}
{De Pontieu}, B., {Hansteen}, V.~H., {Rouppe van der Voort}, L., {van Noort},
  M., \& {Carlsson}, M. 2007{\natexlab{a}}, \apj, 655, 624

\bibitem[{{De Pontieu} {et~al.}(2007{\natexlab{b}}){De Pontieu}, {McIntosh},
  {Hansteen}, {Carlsson}, {Schrijver}, {Tarbell}, {Title}, {Shine}, {Suematsu},
  {Tsuneta}, {Katsukawa}, {Ichimoto}, {Shimizu}, \&
  {Nagata}}]{2007PASJ...59S.655D}
{De Pontieu}, B., {McIntosh}, S., {Hansteen}, V.~H., {et~al.}
  2007{\natexlab{b}}, \pasj, 59, 655

\bibitem[{{De Pontieu} {et~al.}(2007{\natexlab{c}}){De Pontieu}, {McIntosh},
  {Carlsson}, {Hansteen}, {Tarbell}, {Schrijver}, {Title}, {Shine}, {Tsuneta},
  {Katsukawa}, {Ichimoto}, {Suematsu}, {Shimizu}, \&
  {Nagata}}]{2007Sci...318.1574D}
{De Pontieu}, B., {McIntosh}, S.~W., {Carlsson}, M., {et~al.}
  2007{\natexlab{c}}, Science, 318, 1574

\bibitem[{{Fontenla} {et~al.}(1993){Fontenla}, {Avrett}, \&
  {Loeser}}]{1993ApJ...406..319F}
{Fontenla}, J.~M., {Avrett}, E.~H., \& {Loeser}, R. 1993, \apj, 406, 319

\bibitem[{{Giovanelli}(1967)}]{1967AuJPh..20...81G}
{Giovanelli}, R.~G. 1967, Australian Journal of Physics, 20, 81

\bibitem[{{Hansteen} {et~al.}(2006){Hansteen}, {De Pontieu}, {Rouppe van der
  Voort}, {van Noort}, \& {Carlsson}}]{2006ApJ...647L..73H}
{Hansteen}, V.~H., {De Pontieu}, B., {Rouppe van der Voort}, L., {van Noort},
  M., \& {Carlsson}, M. 2006, \apjl, 647, L73

\bibitem[{{Hasan} \& {van Ballegooijen}(2008)}]{2008ApJ...680.1542H}
{Hasan}, S.~S. \& {van Ballegooijen}, A.~A. 2008, \apj, 680, 1542

\bibitem[{{Judge}(2006)}]{2006ASPC..354..259J}
{Judge}, P. 2006, in Astronomical Society of the Pacific Conference Series,
  Vol. 354, Solar MHD Theory and Observations: A High Spatial Resolution
  Perspective, ed. J.~{Leibacher}, R.~F. {Stein}, \& H.~{Uitenbroek}, 259

\bibitem[{{Judge} \& {Peter}(1998)}]{1998SSRv...85..187J}
{Judge}, P.~G. \& {Peter}, H. 1998, Space Science Reviews, 85, 187

\bibitem[{{Krijger} {et~al.}(2001){Krijger}, {Rutten}, {Lites}, {Straus},
  {Shine}, \& {Tarbell}}]{2001A&A...379.1052K}
{Krijger}, J.~M., {Rutten}, R.~J., {Lites}, B.~W., {et~al.} 2001, \aap, 379,
  1052

\bibitem[{{Kurucz}(1979)}]{1979ApJS...40....1K}
{Kurucz}, R.~L. 1979, \apjs, 40, 1

\bibitem[{{Kurucz}(1992{\natexlab{a}})}]{1992RMxAA..23..181K}
{Kurucz}, R.~L. 1992{\natexlab{a}}, Revista Mexicana de Astronomia y
  Astrofisica, vol.~23, 23, 181

\bibitem[{{Kurucz}(1992{\natexlab{b}})}]{1992RMxAA..23..187K}
{Kurucz}, R.~L. 1992{\natexlab{b}}, Revista Mexicana de Astronomia y
  Astrofisica, vol.~23, 23, 187

\bibitem[{{Langangen} {et~al.}(2008){Langangen}, {De Pontieu}, {Carlsson},
  {Hansteen}, {Cauzzi}, \& {Reardon}}]{2008ApJ...679L.167L}
{Langangen}, {\O}., {De Pontieu}, B., {Carlsson}, M., {et~al.} 2008, \apjl,
  679, L167

\bibitem[{{Leenaarts} {et~al.}(2007){Leenaarts}, {Carlsson}, {Hansteen}, \&
  {Rutten}}]{2007A&A...473..625L}
{Leenaarts}, J., {Carlsson}, M., {Hansteen}, V., \& {Rutten}, R.~J. 2007, \aap,
  473, 625

\bibitem[{{Leenaarts} {et~al.}(2006){Leenaarts}, {Rutten}, {S{\"u}tterlin},
  {Carlsson}, \& {Uitenbroek}}]{2006A&A...449.1209L}
{Leenaarts}, J., {Rutten}, R.~J., {S{\"u}tterlin}, P., {Carlsson}, M., \&
  {Uitenbroek}, H. 2006, \aap, 449, 1209

\bibitem[{{Linsky} \& {Avrett}(1970)}]{1970PASP...82..169L}
{Linsky}, J.~L. \& {Avrett}, E.~H. 1970, \pasp, 82, 169

\bibitem[{{Lites} {et~al.}(1993){Lites}, {Rutten}, \&
  {Kalkofen}}]{1993ApJ...414..345L}
{Lites}, B.~W., {Rutten}, R.~J., \& {Kalkofen}, W. 1993, \apj, 414, 345

\bibitem[{{McIntosh} {et~al.}(2008){McIntosh}, {De Pontieu}, \&
  {Tarbell}}]{2008ApJ...673L.219M}
{McIntosh}, S.~W., {De Pontieu}, B., \& {Tarbell}, T.~D. 2008, \apjl, 673, L219

\bibitem[{Neckel(1999)}]{Neckel1999}
Neckel, H. 1999, Sol.\ Phys., 184, 421

\bibitem[{{Reardon} \& {Cavallini}(2008)}]{2008A&A...481..897R}
{Reardon}, K.~P. \& {Cavallini}, F. 2008, \aap, 481, 897

\bibitem[{{Reardon} {et~al.}(2008){Reardon}, {Lepreti},
  {Carbone}, \& {Vecchio}}]{2008ApJ...683L.207R}
{Reardon}, K.~P., {Lepreti}, F., {Carbone}, V., \& {Vecchio}, A.
  2008, \apjl, 683, L207

\bibitem[{{Reardon} {et~al.}(2009){Reardon}, {Uitenbroek}, \&
  {Cauzzi}}]{2008arXiv0810.5260R}
{Reardon}, K.~P., {Uitenbroek}, H., \& {Cauzzi}, G. 2009, A\&A 
in press, arXiv: 0810.5260


\bibitem[{{Rimmele}(2004)}]{2004SPIE.5490...34R}
{Rimmele}, T.~R. 2004, in Society of Photo-Optical Instrumentation Engineers
  (SPIE) Conference Series, Vol. 5490, Society of Photo-Optical Instrumentation
  Engineers (SPIE) Conference Series, ed. D.~{Bonaccini Calia}, B.~L.
  {Ellerbroek}, \& R.~{Ragazzoni}, 34--46

\bibitem[{{Rouppe van der Voort} {et~al.}(2007){Rouppe van der Voort}, {De
  Pontieu}, {Hansteen}, {Carlsson}, \& {van Noort}}]{2007ApJ...660L.169R}
{Rouppe van der Voort}, L.~H.~M., {De Pontieu}, B., {Hansteen}, V.~H.,
  {Carlsson}, M., \& {van Noort}, M. 2007, \apjl, 660, L169

\bibitem[{{Rutten}(2006)}]{2006ASPC..354..276R}
{Rutten}, R.~J. 2006, in Astronomical Society of the Pacific Conference Series,
  Vol. 354, Solar MHD Theory and Observations: A High Spatial Resolution
  Perspective, ed. J.~{Leibacher}, R.~F. {Stein}, \& H.~{Uitenbroek}, 276

\bibitem[{{Rutten}(2007)}]{2007ASPC..368...27R}
{Rutten}, R.~J. 2007, in Astronomical Society of the Pacific Conference Series,
  Vol. 368, The Physics of Chromospheric Plasmas, ed. P.~{Heinzel},
  I.~{Dorotovi{\v c}}, \& R.~J. {Rutten}, 27

\bibitem[{{Rutten} \& {Carlsson}(1994)}]{1994IAUS..154..309R}
{Rutten}, R.~J. \& {Carlsson}, M. 1994, in IAU Symposium, Vol. 154, Infrared
  Solar Physics, ed. D.~M. {Rabin}, J.~T. {Jefferies}, \& C.~{Lindsey}, 309

\bibitem[{{Rutten} \& {Uitenbroek}(1991)}]{1991SoPh..134...15R}
{Rutten}, R.~J. \& {Uitenbroek}, H. 1991, \solphys, 134, 15

\bibitem[{{Rutten} {et~al.}(2008){Rutten}, {van Veelen}, \&
  {S{\"u}tterlin}}]{2008SoPh..251..533R}
{Rutten}, R.~J., {van Veelen}, B., \& {S{\"u}tterlin}, P. 2008, \solphys, 251,
  533

\bibitem[{{Scharmer}(2006)}]{2006A&A...447.1111S}
{Scharmer}, G.~B. 2006, \aap, 447, 1111

\bibitem[{{Tripp} {et~al.}(1978){Tripp}, {Athay}, \&
  {Peterson}}]{1978ApJ...220..314T}
{Tripp}, D.~A., {Athay}, R.~G., \& {Peterson}, V.~L. 1978, \apj, 220, 314

\bibitem[{{Tsiropoula} \& {Schmieder}(1997)}]{1997A&A...324.1183T}
{Tsiropoula}, G. \& {Schmieder}, B. 1997, \aap, 324, 1183

\bibitem[{{Tziotziou}(2007)}]{2007ASPC..368..217T}
{Tziotziou}, K. 2007, in Astronomical Society of the Pacific Conference Series,
  Vol. 368, The Physics of Chromospheric Plasmas, ed. P.~{Heinzel},
  I.~{Dorotovi{\v c}}, \& R.~J. {Rutten}, 217

\bibitem[{{Uitenbroek}(2001)}]{2001ApJ...557..389U}
{Uitenbroek}, H. 2001, \apj, 557, 389

\bibitem[{{van Noort} \& {Rouppe van der Voort}(2006)}]{2006ApJ...648L..67V}
{van Noort}, M.~J. \& {Rouppe van der Voort}, L.~H.~M. 2006, \apjl, 648, L67

\bibitem[{{Vecchio} {et~al.}(2009){Vecchio}, {Cauzzi}, \&
  {Reardon}}]{2009A&A...494..269V}
{Vecchio}, A., {Cauzzi}, G., \& {Reardon}, K.~P. 2009, \aap, 494, 269

\bibitem[{{Vecchio} {et~al.}(2007){Vecchio}, {Cauzzi}, {Reardon}, {Janssen}, \&
  {Rimmele}}]{2007A&A...461L...1V}
{Vecchio}, A., {Cauzzi}, G., {Reardon}, K.~P., {Janssen}, K., \& {Rimmele}, T.
  2007, \aap, 461, L1

\bibitem[{{Vernazza} {et~al.}(1981){Vernazza}, {Avrett}, \&
  {Loeser}}]{1981ApJS...45..635V}
{Vernazza}, J.~E., {Avrett}, E.~H., \& {Loeser}, R. 1981, \apjs, 45, 635

\bibitem[{Wallace {et~al.}(1998)Wallace, Hinkle, \&
  Livingston}]{Wallace+Hinkle+Livingston1998}
Wallace, L., Hinkle, K., \& Livingston, W. 1998, An Atlas of the Spectrum of
  the Solar Photosphere from 13,500 to 28,000~cm$^{-1}$ (3570 to 7405~\AA)
  (National Solar Observatory, Tucson: Technical Report 98-001)

\end{thebibliography}

%%%%%%%%%%%%%%%%%%%%%%%%%%%%%%%%%%%%%%%%%%%%%%%%%%%%%%%%%%%%%%% FIGURE FILE
%% temporary insert in production phase to get labels only
%%%\def\printlabel{}
%%%\def\epsfig#1{\mbox{}\vspace{1cm}\mbox{}}
%%%\def\includegraphics[#1]#2{\mbox{}}
%%%\input{rrmacros2e}
%%%\input{cahafigs}

%%%%%%%%%%%%%%%%%%%%%%%%%%%%%%%%%%%%%%%%%%%%%%%%%%%%%%%%%%%%%%%%%%%%%%% END
\end{document}